\documentclass[12pt, draftclsnofoot, onecolumn]{IEEEtran}
\usepackage[cmex10]{amsmath}
\usepackage{amssymb}
\usepackage{cite}
\usepackage{graphicx}
\usepackage{array,color}
\usepackage{multirow}
\usepackage{amsmath}
\usepackage{stfloats}
\usepackage{graphicx}
\usepackage{subfigure}
\usepackage{tabularx}
\usepackage{epsfig,epsf,color,balance,cite}
\usepackage[caption=false,font=normalsize,labelfont=sf,textfont=sf]{subfig}
\usepackage{setspace}
\usepackage{bm}
\usepackage{textcomp}
\usepackage{algorithmic}
\usepackage{algorithm}
\usepackage{subfigure}
\usepackage{caption}
\usepackage{graphicx}
\usepackage{setspace}
\usepackage{url}
\usepackage{verbatim}
\usepackage{graphicx}
\usepackage{cite}
\newtheorem{theorem}{Theorem}

\newtheorem{lemma}{Lemma}
\begin{document}

\title{Resource Allocation for Uplink Cell-Free Massive MIMO enabled URLLC in a Smart Factory}

\author{Qihao Peng, ~\IEEEmembership{Student Member,~IEEE}, Hong Ren, ~\IEEEmembership{Member,~IEEE}, \\ Cunhua Pan, ~\IEEEmembership{Member,~IEEE}, 
Nan Liu, ~\IEEEmembership{Member,~IEEE}, and Maged Elkashlan, ~\IEEEmembership{Senior Member,~IEEE}
\thanks{This works was supported in part by the National Key Research and Development Project (Grant No. 2019YFE0123600), National Natural Science Foundation of China (Grant No. 62101128), and Basic Research Project of Jiangsu Provincial Department of Science and Technology (Grant No. BK20210205). The work of Qihao Peng was supported by the China Scholarship Council. (\emph{Corresponding author: Hong Ren and Cunhua Pan}.)}
\thanks{Q. Peng and M. Elkashlan are with the School of Electronic Engineering and Computer Science at Queen Mary University of London, U.K. (e-mail: \{q.peng, maged.elkashlan\}@qmul.ac.uk). H. Ren, C. Pan, and N. Liu are with National Mobile Communications Research Laboratory, Southeast University, Nanjing, China. (e-mail:\{hren,cpan,nanliu\}@seu.edu.cn).}
}

\maketitle
\vspace{-0.3cm}
\begin{abstract}
Smart factories need to support the simultaneous communication of multiple industrial Internet-of-Things (IIoT) devices with ultra-reliability and low-latency communication (URLLC). Meanwhile, short packet transmission for IIoT applications incurs performance loss compared to traditional long packet transmission for human-to-human communications. On the other hand, cell-free massive multiple-input and multiple-output (CF mMIMO) technology can provide uniform services for all devices by deploying distributed access points (APs). In this paper, we adopt CF mMIMO to support URLLC in a smart factory. Specifically, we first derive the lower bound (LB) on achievable uplink data rate under the finite blocklength (FBL) with imperfect channel state information (CSI) for both maximum-ratio combining (MRC) and full-pilot zero-forcing (FZF) decoders. \textcolor{black}{The derived LB rates based on the MRC case have the same trends as the ergodic rate, while LB rates using the FZF decoder tightly match the ergodic rates}, which means that resource allocation can be performed based on the LB data rate rather the exact ergodic data rate under FBL. The \textcolor{black}{log-function method} and successive convex approximation (SCA) are then used to approximately transform the non-convex weighted sum rate problem into a series of geometric program (GP) problems, and an iterative algorithm is proposed to jointly optimize the pilot and payload power allocation. Simulation results demonstrate that CF mMIMO significantly improves the average weighted sum rate (AWSR) compared to centralized mMIMO. An interesting observation is that increasing the number of devices improves the AWSR for CF mMIMO whilst the AWSR remains relatively constant for centralized mMIMO.
\end{abstract}


\section{Introduction}

The smart factory is envisioned as one of the most fundamental application scenarios in the next generation of industrial systems, which entails ultra-reliable and low-latency communication (URLLC) for wireless connected terminals \cite{ref1,ref2}. For typical industrial applications, \textcolor{black}{wireless packet has several hundred bits and is delivered with high reliability (i.e., above $1-10^{-6}$) and low latency (i.e., below 1 ms) to fulfill the goal of real-time and precise control \cite{ref1}, and thus the channel blocklength is finite. According to Shannon coding theorem, the decoding error probability (DEP) always approaches zero when the channel blocklength is infinity \cite{1948Shanon}. However, the DEP cannot approach zero when the channel blocklength is limited \cite{popovski2019wireless,ref3}, which cannot be ignored in the transmission design}. 


Recently, the achievable data rate in terms of the finite blocklength (FBL), the DEP and signal-to-noise ratio (SNR) was derived in \cite{ref3}. Since then, significant efforts have been devoted to the transmission design based on the capacity under FBL \cite{ref5,Hong-twc,cao2019joint,ref6,ref7,Hong-tcom,monemi2020performance}. Multiple messages were grouped into a single packet to reduce the transmission latency \cite{ref5}. The authors \textcolor{black}{of} \cite{Hong-twc} jointly optimized the power and blocklength for two devices under four schemes, namely, orthogonal multiple access (OMA), non-orthogonal multiple access (NOMA), relay, and cooperative relay. \textcolor{black}{Similarly, the blocklength was optimized to satisfy the stringent requirements of the DEP and latency \cite{cao2019joint}.} The DEP of relay-assisted transmission under FBL was analyzed in \cite{ref6} with perfect channel state information (CSI), which was further extended to the imperfect CSI case in \cite{ref7}. Resource allocation for a secure URLLC scenario was studied in \cite{Hong-tcom}. \textcolor{black}{Besides, the unmanned aerial vehicle (UAV) was deployed to deal with the blockage issue in \cite{monemi2020performance} under FBL.}  

To support multiple devices simultaneously, massive multiple-input and multiple-output (mMIMO) for URLLC \cite{ref13,ref13b1,ref13a,ref13b} has attracted extensive research attention due to its appealing feature of a large number of spatial degrees of freedom \cite{ref11,you2020shannon}. In addition, the channel hardening effect of mMIMO fits well with the rich scattering environment in smart factories \cite{li2017review}. The authors \textcolor{black}{of} \cite{ref13} and \cite{ref12} investigated the network availability and analyzed the relationship between the DEP and the number of antennas. The system performance with severe shadow fading was studied under the stringent requirements on URLLC \cite{ref13a}. Moreover, the closed-form expression of the average secrecy throughput was derived in \cite{ref13b}. Besides, joint pilot and payload transmission power allocation for mMIMO URLLC was studied in \cite{ref13d}, where the best local approximation and geometric program (GP) were introduced to solve the optimization problem. Although adopting mMIMO can provide enhanced service for multiple devices, it still faces some critical challenges in the centralized mMIMO-enabled smart factory scenarios, e.g., i) {\emph{the poor quality of service (QoS) for the devices far away from the AP}}; and ii) {\emph{the blockage issue between the AP and the devices}}.

\textcolor{black}{To tackle the above challenges, the promising technique named cell free mMIMO (CF mMIMO) has been proposed \cite{interdonato2019ubiquitous}. Unlike typical cell-centric networks, CF mMIMO can support user-centric transmissions, where all access points (APs) jointly serve all devices without cell boundaries \cite{ref16}.} Therefore, CF mMIMO is regarded as the future paradigm for the next generation of wireless communication \cite{ref14}. For a distributed precoding scheme, Giovanni {\it{et al.}} developed a full-pilot zero-forcing (FZF) method with orthogonal pilots to suppress inter-cell interference, and then proposed a local partial zero-forcing precoding for reusing pilots \cite{ref18}. The centralized minimum mean-square error (MMSE) processing was provided in \cite{ref18a}. To reduce the implementation complexity, the user-centric approach was adopted in \cite{ref18b}. Besides, the impact of APs' density on the system performance was investigated in \cite{ref18c}. The authors \textcolor{black}{of} \cite{ref19} considered the power allocation problem, where each AP is equipped with a single antenna.

Due to the appealing features of CF mMIMO, some researchers have already noticed the advantages of adopting CF mMIMO to support multiple devices with URLLC. Specifically, the network availability in terms of the DEP was analyzed in \cite{CF21Per}, which demonstrated the performance gains over the centralized mMIMO. The authors \textcolor{black}{of} \cite{ref20} considered two power allocation problems with the objectives of maximizing the minimum data rate and maximizing the energy efficiency, where the simple conjugate beamforming was adopted at the \textcolor{black}{single-antenna} APs. \textcolor{black}{However, it has been shown that single-antenna APs can hardly improve the reliability of estimated channels (i.e., channel hardening), unless ultra-high-density APs are deployed, which is theoretically possible but practically unrealistic \cite{2018Channelharden}.} To the best of our knowledge, we are the first to investigate the URLLC enabled by CF mMIMO where each AP is equipped with multiple antennas. \textcolor{black}{In addition, we jointly optimize the pilot power and payload power allocation to maximize the weighted sum rate. Different from the previous works relying on GP in \cite{van2018joint,ref13d} or sum rate optimization relying on weighted MMSE in \cite{van2020power}, jointly allocating power based on the FBL is more challenging in CF mMIMO systems.} Our contributions are summarized as follows.

\begin{enumerate}
  \item \textcolor{black}{The lower bounds (LBs) on the achievable uplink data rates under FBL for the maximum-ratio combining (MRC) and FZF schemes are derived for the CF mMIMO. Simulation results confirm that there exists a gap between the LB rate based on the MRC scheme and the ergodic data rate, and the LB rate using the FZF decoder can tightly match the ergodic rate, which provides tractable expressions for power allocation.}  
  \item For the MRC decoder, due to the non-convex weighted sum rate expression, it is challenging to obtain the optimal solution. \textcolor{black}{The log-function method is adopted to approximate the objective function in an iterative manner.} Meanwhile, the numerator of the signal-to-interference-plus-noise (SINR) is a posynomial function and cannot be transformed into a GP problem. To tackle this issue, the successive convex approximation (SCA) is adopted to approximately transform the numerator into a series of monomial functions, and then the optimization problem can be readily solved by CVX.
  \item For the FZF decoder, the expression of the SINR is more complicate than that \textcolor{black}{of} the MRC decoder, and thus it is more challenging to solve the optimization problem. Different from the MRC case where the numerator of the SINR contains only one pilot power allocation variable, the numerator of the SINR based on the FZF decoder contains all devices' pilot power allocation variables. We first prove that the numerator of SINR is a convex function by examining its Hessian matrix, based on which we derive its LB by using Jensen's inequality. Fortunately, the LB of the numerator of the SINR is a monomial function, and the original optimization problem can be approximately addressed by solving a series of GP problems. Finally, an iterative algorithm is proposed to jointly optimize the pilot and payload power allocation.
  \item Simulation results demonstrate the rapid convergence of our proposed algorithms, and also validate that our proposed method has a remarkable performance improvement over the benchmark schemes.
\end{enumerate}

The remainder of this paper is organized as follows. In Section II, the system model is provided, and then the LB date rate under FBL based on statistical CSI are derived for the MRC and FZF decoders, respectively. In Section III, the optimization problem of maximizing weighted sum rate is simplified into a GP problem and an iterative algorithm is proposed by jointly optimizing the pilot and payload power for the MRC decoder. The iterative algorithm for jointly allocating power for the FZF case is given in Section IV. Then, simulation results and analysis are presented in Section V. Finally, the conclusion is drawn in Section VI.

\section{System Model and Problem Formulation}
\subsection{System Model}

\begin{figure}[t]
\centering
\includegraphics[width=3.2in]{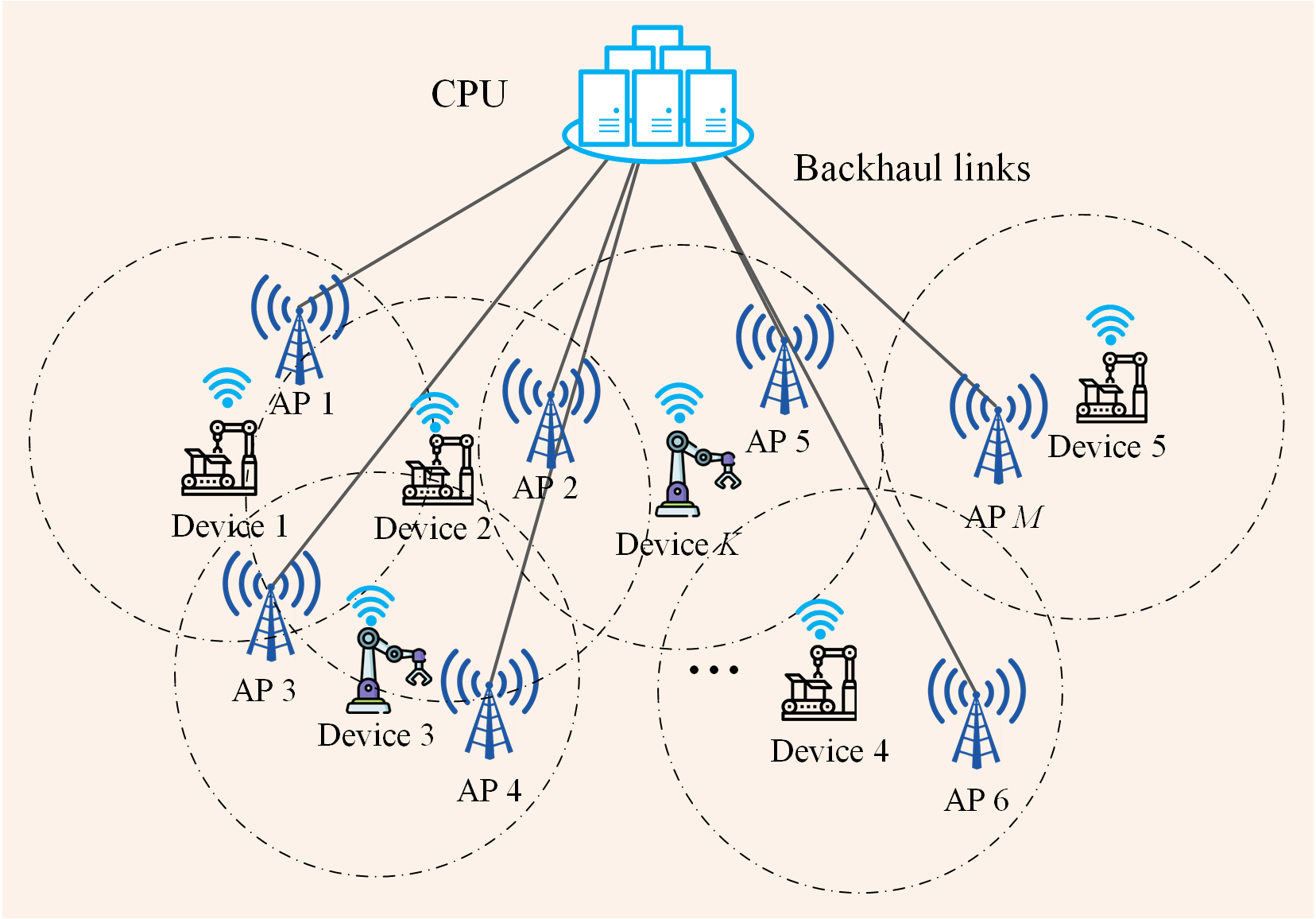}
\caption{Smart factory scenario where CF mMIMO serves multiple devices.}
\label{fig:system model}
\end{figure}

We consider an uplink CF mMIMO-enabled smart factory illustrated in Fig. \ref{fig:system model}, where each AP and each device are equipped with $N$ antennas and a single antenna, respectively. These APs are connected to a central processing unit (CPU) through backhauls. The channel vector ${\bf{g}}_{m,k} \in {\mathbb{C}}^{N \times 1} $ between the $m$th AP and the $k$th device is modeled as
\begin{equation}
\setlength\abovedisplayskip{5pt}
\setlength\belowdisplayskip{5pt}
\label{cofficient_g}
{{\bf{g}}_{m,k}} = \sqrt {{\beta _{m,k}}} {{\bf{h}}_{m,k}},
\end{equation}
where $\beta _{m,k}$ is the large-scale fading and ${\bf{h}}_{m,k} \in \mathcal{CN} \left( {{\bf{0}},{\bf{I}}_N} \right)$ represents the small-scale fading. For simplicity, we denote ${{\bf{G}}_m} = [{\bf{g}}_{m,1},{\bf{g}}_{m,2},\cdot  \cdot  \cdot,{\bf{g}}_{m,K}]$ as the channel matrix from all the devices to the $m$th AP.

The received signal at the $m$th AP is given by
\begin{equation}
\label{received_vector}
\setlength\abovedisplayskip{5pt}
\setlength\belowdisplayskip{5pt}
{{\bf{y}}_m} = {{\bf{G}}_m}\sqrt {{{\bf{P}}^d}} {\bf{s}} + {{\bf{n}}_m},
\end{equation}
where ${\bf{P}}^d = \text{diag}\left\{ {{p_1^d,p_2^d, \cdot  \cdot  \cdot ,p_K^d} } \right\}$ is the payload power of $K$ devices, \textcolor{black}{$p_k^d$ is the $k$th device's transmission power}, ${\bf{s}} \in  {{\mathbb{C}}^{K \times 1}}$ is the transmission symbol vector with zero mean and unit covariance matrix, and ${\bf{n}}_m$ is the normalized noise vector following the distribution of $\mathcal{CN} \left( {{\bf{0}},{\bf{I}}_N} \right)$. \textcolor{black}{Then, the $m$th AP decodes the received signal based on the locally estimated channel, and then delivers the decoded signal to the CPU \cite{ref16}. Finally, the CPU combines the signals to obtain the information based on the user-centric approach.}


\subsection{Channel Estimation}
It is assumed that each AP needs to estimate the CSI from all the devices based on TDD protocol. In order to distinguish the channels from different devices, $K$ devices are allocated with orthogonal pilot sequences. Let us define $L_p$ ($L_p = K$) as the length of the pilot sequence for each device and \textcolor{black}{${\bf{q}}_k \in  {{\mathbb{C}}^{K \times 1}} $} as the pilot sequence of the $k$th device, $\forall k$. The finite blocklength $L$ is divided into blocklength $L_p$ for pilot sequence and blocklength $(L - L_p)$ for data transmission, respectively. Assume that the bandwidth is $B$. Then, the time durations for channel estimation and data transmission are $t_p = L_p / B$ and $t_d = L_d/B$, respectively.

In the training phase, $K$ orthogonal pilot sequences are received by all APs, and then the received pilot signal at \textcolor{black}{the} $m$th AP is denoted as
\begin{equation}
\label{received_pilot}
\setlength\abovedisplayskip{5pt}
\setlength\belowdisplayskip{5pt}
{{\bf{Y}}^p_{m}} = \sum\limits_{k = 1}^K {{{\bf{g}}_{m,k}}\sqrt {{K}p_k^p} {\bf{ q }}_k^H}  + {{\bf{N}}^p_{m}},
\end{equation}
where $p_k^p$ is the pilot power of the $k$th device, and \textcolor{black}{${{\bf{N}}_{m}^p} \in  {{\mathbb{C}}^{N \times K}}$} is the additive Gaussian noise matrix at the $m$th AP, each element of which is independent and follows the distribution of $\mathcal{CN} \left( {{{0}},{{1}}} \right)$. By multiplying (\ref{received_pilot}) with ${\bf{ q }}_k$, we have
\begin{equation}
\label{received_channel}
\setlength\abovedisplayskip{5pt}
\setlength\belowdisplayskip{5pt}
{{\bf{\hat y}}_{m,k}^p} = \frac{1}{{\sqrt {Kp_k^p} }}{{\bf{Y}}^p_{m}}{{\bf{q}}_k} = {{\bf{g}}_{m,k}} + {\bf{n}}_{m,k}^p,
\end{equation}
where ${\bf{n}}_{m,k}^p = \frac{1}{{\sqrt {Kp_k^p} }}{{\bf{N}}^p_{m}}{{\bf{q}}_k}$. Based on (\ref{received_channel}), the MMSE estimate for ${\bf{g}}_{m,k}$ is
\begin{equation}
\label{estimate_gmk}
\setlength\abovedisplayskip{5pt}
\setlength\belowdisplayskip{5pt}
{{{\bf{\hat g}}}_{m,k}} = \frac{{K{p_{k}^p}{\beta _{m,k}}}}{{K{p_{k}^p}{\beta _{m,k}} + 1}}{{{\bf{\hat y}}}^p_{m,k}},
\end{equation}
which follows the distribution of $\mathcal{CN} \left( {{{\bf {0}}},{{{\lambda_{m,k}}{{\bf{I}}_N}}}} \right)$, and $\lambda_{m,k}$ is given by
\begin{equation}
\label{gama_mk}
\setlength\abovedisplayskip{5pt}
\setlength\belowdisplayskip{5pt}
{\lambda_{m,k}} = \frac{{Kp_k^p{{\left( {{\beta _{m,k}}} \right)}^2}}}{{Kp_k^p{\beta _{m,k}} + 1}}.
\end{equation}
Then, let us denote ${{{\bf{\tilde g}}}_{m,k}} = {{\bf{g}}_{m,k}} - {{{\bf{\hat g}}}_{m,k}}$ as the estimation error, which is independent of $ {{{\bf{\hat g}}}_{m,k}}$ and follows the distribution of $\mathcal{CN} \left( {{{\bf{0}}},{{ \left( {{\beta _{m,k}} - {\lambda _{m,k}}} \right){{\bf{I}}_N}}}} \right)$ .

\subsection{Achievable Date Rate under Finite Blocklength}
As previously stated, \textcolor{black}{Shannon capacity under infinite channel blocklength is no longer applicable due to FBL}. \textcolor{black}{Based on the result in \cite{ref13d,scarlett2016dispersion}, the interference can be treated as Gaussian noise. Therefore, the achievable data rate can be approximated as} 
\begin{equation}
\label{urllc_rate}
\setlength\abovedisplayskip{5pt}
\setlength\belowdisplayskip{5pt}
{R_k} \approx B \left[\left( {1 - \eta } \right){\log _2}\left( {1 + \gamma_k} \right) - \sqrt {\frac{{\left( {1 - \eta } \right){V_k\left(\gamma_k\right)}}}{L}} \frac{{{Q^{ - 1}}\left( {{\varepsilon _k}} \right)}}{{\ln 2}}\right],
\end{equation}
where $\eta = K / L$, ${\gamma}_k$ is the $k$th device's SINR, ${{\varepsilon _k}}$ is DEP,  $V_k$ is the channel dispersion with  ${V_k\left(\gamma_k\right)} = 1 - {\left( {1 + {\gamma_k}} \right)^{ - 2}}$, and ${Q^{ - 1}}\left( {{\varepsilon _k}} \right)$ is the inverse function of $Q\left( {{\varepsilon _k}} \right) = \frac{1}{{\sqrt {2\pi } }}\int_{{\varepsilon _k}}^\infty  {{{\rm{e}}^{{{ - {t^2}} \mathord{\left/
 {\vphantom {{ - {t^2}} 2}} \right.
 \kern-\nulldelimiterspace} 2}}}{\rm{d}}t}$ of the $k$th device, $\forall k$.

\textcolor{black}{For the $m$th AP, the linear vector ${\bf{a}}_{m,k}$ for the $k$th device is based on the locally estimated channel in (\ref{estimate_gmk}), which is given by \cite{ref16,ref18}}
\begin{equation}
\setlength\abovedisplayskip{5pt}
\setlength\belowdisplayskip{5pt}
\label{linear_decoder}
{{\bf{a}}_{m,k}} = \left\{ {\begin{array}{*{20}{c}}
{{{{\bf{\hat g}}}_{m,k}},}&{{\rm{MRC}}},\\
{\frac{{{{{\bf{\hat G}}}_m}{{\left( {{{{\bf{\hat G}}}_m^H}{{{\bf{\hat G}}}_m}} \right)}^{ - 1}}{{\bf{e}}_k}}} {\sqrt{{{\mathbb{E}}{{{\left\| {{{{\bf{\hat G}}}_m}{{\left[ {{\bf{\hat G}}_m^H{{{\bf{\hat G}}}_m}} \right]}^{ - 1}}{{\bf{e}}_k}} \right\|}^2}}}}},}&{{\rm{FZF}}},
\end{array}} \right.
\end{equation}
where $\mathbb{E}\left \{ \cdot \right \}$ denotes the expectation operator, ${\bf{\hat G}}_m = \left[ {{{{\bf{\hat g}}}_{m,1}},{{{\bf{\hat g}}}_{m,2}}, \cdot  \cdot  \cdot ,{{{\bf{\hat g}}}_{m,K}}} \right]$ is the estimated channel matrix between all the devices and the $m$th AP, and ${\bf{e}}_{k}$ represents the $k$th column of ${\bf {I}}_K$. \textcolor{black}{As mentioned before, the $m$th AP multiplies the received signal with the decoding vector ${\bf{a}}_{m,k}$ to obtain the $k$th device's information, which can be denoted by} 
\begin{equation}
	\setlength\abovedisplayskip{5pt}
	\setlength\belowdisplayskip{5pt}
	{{{y}}_{m,k}^d} = {\bf{a}}_{m,k}^{H}{{\bf{G}}_m}\sqrt {{{\bf{P}}^d}} {\bf{s}} + {\bf{a}}_{m,k}^{H}{{\bf{n}}_m}.
\end{equation}
\textcolor{black}{Besides, to reduce the effect of small-scale fading, we assume that each AP treats the mean of the effect channel gain as the true channel for signal detection \cite{ref16,ref14,ref18}. Then, each AP conveys the decoded signal to the CPU. Based on the user-centric approach, the CPU can combine signals from some APs to decode the $k$th device's information, and we denote ${\mathcal{M}}_k$ as the set of APs that serve the $k$th device for ease of exposition.} \textcolor{black}{Therefore, the CPU combines the signals from APs in the set of ${\mathcal{M}}_k$ to acquire the $k$th device's information, which is given by}
\begin{align}
\setlength\abovedisplayskip{5pt}
\setlength\belowdisplayskip{5pt}
y_k^d  &= \mathbb{E} \underbrace {\left\{ {\sum\limits_{m \in {\mathcal{M}}_k} {{{\left( {{\bf{ a}}_{m,k}} \right)}^H}{\bf{g}}_{m,k}\sqrt {{p_k^d}} } } \right\}}_{{\rm{DS}}_k}{s_k}  \notag \\ 
&\!+\!\underbrace {\sqrt {{p_k^d}} \!\left\{\!{\sum\limits_{m \in {\mathcal{M}}_k}\!\!\! {{{\left( {{\bf{ a}}_{m,k}} \right)}^H}{\bf{g}}_{m,k}}  \!-\! \mathbb{E}{\left\{\! {\sum\limits_{m \in {\mathcal{M}}_k}\!\!\! {{{\left( {{\bf{ a}}_{m,k}} \right)}^H}{\bf{g}}_{m,k} } } \!\right\}}  } \!\right\}}_{{\rm {LS}}_k}\!{s_k}, \label{kth_statistical_signal} \\
 &\! + \!\sum\limits_{k' \ne k}^K \! { \underbrace {\sum\limits_{m \in {\mathcal{M}}_k}\!\!{{{\left( {{\bf{ a}}_{m,k}} \right)}^H}{\bf{g}}_{m,k'}\sqrt {{p_{k'}^d}} } }_{{\rm {UI}}_{k,k'}}\!{s_{k'}}}\! +\!\!\!\! \underbrace {\sum\limits_{m \in {\mathcal{M}}_k}\!\! {{{\left( {{\bf{ a}}_{m,k}} \right)}^H}{\bf{n}}_m} }_{{{\rm {N}}_k}} \notag,
\end{align}
where \textcolor{black}{$p_k^d$ is the $k$th device's transmission power,} $s_k$ is the transmitted information, ${\bf{n}}_m$ is the noise vector, ${\rm{DS}}_k$ is the desired signal, ${\rm {LS}}_k$ is the leaked signal, ${\rm {UI}}_{k,k'}$ represents the interference of the $k'$th device, and ${\rm {N}}_k$ is the noise term. Then, the SINR at the $k$th device is given by
\begin{equation}
\setlength\abovedisplayskip{5pt}
\setlength\belowdisplayskip{5pt}
\label{kth_SINR}
\gamma _k = \frac{{{{\left| {{\rm{DS}}_{k}} \right|}^2}}}{{{{\left| {{\rm{LS}}_{k}} \right|}^2} + \sum\nolimits_{k' \ne k}^K {{{\left| {{\rm{UI}}_{k,k'}} \right|}^2}}  + {{\left| {{{\rm{N}}_k}} \right|}^2}}}.
\end{equation}

\textcolor{black}{Due to the channel hardening, the impact of random channel gain on communication is negligible. Therefore}, we consider the optimization of pilot power and payload power \textcolor{black}{that are} only based on the large-scale CSI, which varies much slower than the instantaneous CSI. In this case, the pilot and payload power only needs to be updated once the large-scale CSI changes rather than the rapidly varying instantaneous CSI. As a result, we first need to derive the ergodic data rate of the devices. The ergodic capacity of the $k$th device under FBL is given by
\begin{equation}
\setlength\abovedisplayskip{5pt}
\setlength\belowdisplayskip{5pt}
\label{rw_rate}
\begin{split}
{\bar R_k} &\approx \mathbb{E} \left\{ \!\! B \frac{{1 - \eta }}{{\ln 2}} \!\! \left[\! {\ln \left( {1 \!+\! {\gamma _k}} \right) \!\!- \!\!\frac{{{Q^{ - 1}}\left( {{\varepsilon _k}} \right)}}{{\sqrt {L\left( {1 - \eta } \right)} }}\!\sqrt {\frac{{\frac{2}{{{\gamma _k}}} + 1}}{{{{\left( {\frac{1}{{{\gamma _k}}} + 1} \right)}^2}}}} } \!\!\right] \right\}, \\
&\triangleq B \frac{{1 - \eta }}{{\ln 2}} \mathbb{E} \left\{ f_k \left( \frac{1}{\gamma_k} \right)\right\},
\end{split}
\end{equation}
\textcolor{black}{where $f_k \left( x \right) = \ln(1+ \frac{1}{x}) + \frac{{{Q^{ - 1}}\left( {{\varepsilon _k}} \right)}}{{\sqrt {L\left( {1 - \eta } \right)} }}\sqrt {\frac{{2x+ 1}}{{{{\left( {x + 1} \right)}^2}}}}$ denotes a function of the $k$th device's DEP $\varepsilon _k$.}

However, it is challenging to derive the closed-form expression of the ergodic data rate. Instead, we aim to derive its LB. To this end, we first provide the following results.
Since the rate $R_k$ is no smaller than $0$, the following inequality holds
\begin{equation}
\setlength\abovedisplayskip{5pt}
\setlength\belowdisplayskip{5pt}
\label{a_region}
\frac{{{Q^{ - 1}}\left( {{\varepsilon _k}} \right)}}{{\sqrt {L\left( {1 - \eta } \right)} }} \le \frac{{\left( {\frac{1}{{{\gamma _k}}} + 1} \right)\ln \left( {1 + {\gamma _k}} \right)}}{{\sqrt {\frac{2}{{{\gamma _k}}} + 1} }} \buildrel \Delta \over =  g\left( \frac{1}{{\gamma}_k} \right ).
\end{equation}

As the first-order derivative of $g\left( x \right )$ is less than $0$, $g\left( x \right )$ is monotonically decreasing. Besides, the feasible region of $f_{k} \left(x \right)$ is $0 \le x \le {g^{ - 1}}\left( {\frac{{{Q^{ - 1}}\left( {{\varepsilon _k}} \right)}}{{\sqrt {L\left( {1 - \eta } \right)} }}} \right)$. Then, we have the following lemma.

\begin{lemma}
\label{x_region}
Function $f_{k} \left(x \right)$ is decreasing and convex when $ 0 < x \le {g^{ - 1}}\left( {\frac{{{Q^{ - 1}}\left( {{\varepsilon _k}} \right)}}{{\sqrt {L\left( {1 - \eta } \right)} }}} \right) $.

\emph{Proof}: Please refer to Appendix B in \cite{ref20a}. $\hfill\blacksquare$

\end{lemma}

By using Jensen's inequality and Lemma \ref{x_region}, the ergodic data rate is lower bounded by
\begin{equation}
\setlength\abovedisplayskip{5pt}
\setlength\belowdisplayskip{5pt}
\label{rate}
{\bar R_k} \ge {\hat R}_k \triangleq B \frac{{1 - \eta }}{{\ln 2}} {f_k}\left( {{1 \mathord{\left/
 {\vphantom {1 {{{\hat \gamma }_k}}}} \right.
 \kern-\nulldelimiterspace} {{{\hat \gamma }_k}}}} \right),
\end{equation}
where ${\hat R}_k$ is the LB of the $k$th device's ergodic data rate, and ${{\hat \gamma }_k}$ is ${{\hat \gamma }_k} = \frac{1}{\mathbb{E} \left( {{1 \mathord{\left/
 {\vphantom {1 {{{\hat \gamma }_k}}}} \right.
 \kern-\nulldelimiterspace} {{{ \gamma }_k}}}} \right)}$.

In the following, we derive the expression of ${\hat R}_k$ for the MRC and FZF decoders, respectively. \textcolor{black}{Specifically}, we have extended the results for the centralized mMIMO in \cite{ref13d} to the more general user-centric CF mMIMO.

\begin{theorem}
\label{MRC_SINR_T}
The ergodic achievable rate for the $k$th device \textcolor{black}{using} the MRC decoder with FBL can be lower bounded by
\begin{equation}
\setlength\abovedisplayskip{5pt}
\setlength\belowdisplayskip{5pt}
\label{MRC_LB_rate}
 {\hat R}_k^{\rm MRC} \triangleq  B \frac{{1 - \eta }}{{\ln 2}} f_k \left( \frac{1}{{\hat \gamma }_k^{\rm MRC}} \right),
\end{equation}
where ${\hat \gamma }_k^{\rm MRC}$ is denoted as
\begin{equation}
\setlength\abovedisplayskip{5pt}
\setlength\belowdisplayskip{5pt}
\label{MRC_SINR_LB}
\hat \gamma _k^{{\rm{MRC}}} = \frac{{Np_k^d{{\left( {\sum\limits_{m \in {\mathcal{M}}_k} {{\lambda _{m,k}}} } \right)}^2}}}{{\sum\limits_{k' = 1}^K {\sum\limits_{m \in {\mathcal{M}}_k}{p_{k'}^d{\lambda _{m,k}}{\beta _{m,k'}}} }  + \sum\limits_{m \in {\mathcal{M}}_k}{{\lambda _{m,k}}} }}.
\end{equation}

\emph{Proof}: Please refer to Appendix \ref{MRC_SINR_P}. $\hfill\blacksquare$

\end{theorem}

\begin{theorem}
\label{FZF_SINR_T}
The $k$th device's ergodic achievable rate for the FZF decoder is lower bounded by
\begin{equation}
\setlength\abovedisplayskip{5pt}
\setlength\belowdisplayskip{5pt}
\label{FZF_LB_rate}
 {\hat R}_k^{\rm FZF} \triangleq  B \frac{{1 - \eta }}{{\ln 2}} f_k \left( \frac{1}{{\hat \gamma }_k^{\rm FZF}} \right),
\end{equation}
where ${\hat \gamma }_k^{\rm FZF}$ is denoted as
\begin{equation}
\setlength\abovedisplayskip{5pt}
\setlength\belowdisplayskip{5pt}
\label{SINR_ZF}
\hat \gamma _k^{{\rm{FZF}}} = \frac{{{p_k^d}\left( {N - K} \right){{\left( {\sum\limits_{m \in {\mathcal{M}}_k} {\sqrt {{\lambda _{m,k}}} } } \right)}^2}}}{{|{\cal M}_k| + \sum\limits_{k' = 1}^K {{p_{k'}^d}\sum\limits_{m \in {\mathcal{M}}_k} {\left( {{\beta _{m,k'}} - {\lambda _{m,k'}}} \right)} } }},
\end{equation}
where $|{\cal M}_k|$ means the cardinality of the set ${\cal M}_k$, and the number of antennas $N$ should be larger than the number of devices $K$.

\emph{Proof}: Please refer to Appendix \ref{FZF_SINR_P}. $\hfill\blacksquare$

\end{theorem}

\subsection{Problem Formulation}
\textcolor{black}{In smart factory, the weight sum rate is an effective method to satisfy various devices' requirements \cite{ref13d,van2020power}}. Therefore, we aim to jointly optimize the pilot power and payload power to maximize the weighted sum rate of all devices subject to the data rate constraints and the total energy constraints. Mathematically, the optimization problem can be formulated as
\begin{subequations}
\setlength\abovedisplayskip{5pt}
\setlength\belowdisplayskip{5pt}
\label{MRC_optimization}
\begin{align}
\mathop {\max }\limits_{\left\{ {p_k^p} \right\},\left\{ {p_k^d} \right\}} & \sum\limits_{k = 1}^K {{w_k}{{\hat R}_k}} \label{MRC_optimization_a}\\
{\rm{s}}{\rm{.t}}{\rm{.}}\;\;\;\; & {{\hat R}_k} \ge R_k^{{\rm{req}}},\forall k,  \label{MRC_optimization_b}\\
& {Kp_k^p + \left( {L - K} \right)p_k^d \le {E_k},\forall k}, \label{MRC_optimization_c}
\end{align}
\end{subequations}
where ${{\hat R}_k}$ denotes the LB data rate for either MRC or FZF, $w_k$ is the weight of the $k$th device, constraint (\ref{MRC_optimization_b}) denotes the minimum data rate requirements of the devices, and constraint (\ref{MRC_optimization_c}) means that the energy consumption of each device is limited.

\textcolor{black}{Unlike the Shannon Capacity based on infinite blocklength (i.e., the max-min fairness problem in \cite{van2018joint} or sum rate in \cite{van2020power}), the expression of data rate under FBL is more complicated. In addition, the weighted sum rate problem is an NP-hard problem and is more challenging to solve under imperfect CSI and FBL.} To deal with this difficulty, we introduce some methods for simplifying the problem, and then solve the problem with polynomial-time complexity.

Using Lemma \ref{x_region}, the constraint (\ref{MRC_optimization_b}) can be simplified into the $k$th device's minimal SINR requirement, denoted as
\begin{equation}
\setlength\abovedisplayskip{5pt}
\setlength\belowdisplayskip{5pt}
\label{transformation}
\hat \gamma _k \ge \frac{1}{{f_k^{ - 1}\left( {\frac{{R_k^{{\rm{req}}}\ln 2}}{{(1 - \eta)B }}} \right)}},
\end{equation}
where $\hat \gamma _k$ represents the SINR \textcolor{black}{of the} $k$th device by using MRC or FZF decoder, respectively. Then, the auxiliary variables ${\chi _k}$ is introduced to equivalently transform (\ref{MRC_optimization}) into the following optimization problem
\begin{subequations}
\setlength\abovedisplayskip{5pt}
\setlength\belowdisplayskip{5pt}
\label{MRC_optimization_trans}
\begin{align}
\mathop {\max }\limits_{\left\{ {p_k^p} \right\},\left\{ {p_k^d} \right\},\left\{ {\chi_k} \right\}} & {\sum\limits_{k = 1}^K {{w_k B}\frac{{\left( {1 - \eta } \right)}}{{\ln 2}}\left[ {\ln \left( {1 + {\chi _k}} \right) - {\alpha _k}G\left( {{\chi _k}} \right)} \right]} } \label{MRC_optimization_trans_a}\\
\text{s.t.} \;\;\;\;\; & \hat \gamma _k \ge {\chi _k},\forall k,  \label{MRC_optimization_trans_b}\\
&{\chi _k} \ge \frac{1}{{f_k^{ - 1}\left( {\frac{{R_k^{{\rm{req}}}\ln 2}}{{(1 - \eta)B }}} \right)}},\forall k, \label{MRC_optimization_trans_c} \\
&{\rm{ \left(\ref{MRC_optimization_c}\right)}}, \label{MRC_optimization_trans_d}
\end{align}
\end{subequations}
where $G\left( {{\chi _k}} \right)$ is defined as $G\left( {{\chi _k}} \right) \triangleq \sqrt {\frac{{\frac{2}{{{\chi _k}}} + 1}}{{{{\left( {\frac{1}{{{\chi _k}}} + 1} \right)}^2}}}}$, and $\alpha_k$ is $\alpha_k = {\frac{{{Q^{ - 1}}\left( {{\varepsilon _k}} \right)}}{{\sqrt {L\left( {1 - \eta } \right)} }}}$. Due to the different expressions of $\hat \gamma _k $, we provide solutions for the power allocation for the MRC and FZF decoders, respectively.
\section{Power Allocation for the Case of MRC}
In this section, we aim to solve the weighted sum rate maximization problem for the MRC decoder.

\subsection{Joint Optimization}
As seen in (\ref{MRC_optimization_trans_a}), it it challenging to solve the optimization problem due to the complicated functions of $\ln \left( 1 + \chi_k \right)$ and $G\left(\chi_k\right)$. To simplify the objective function, two lemmas are introduced in the following.

\begin{lemma}
\label{lnx}
For any given ${\hat {x}} \ge 0$, function $\ln\left(1 + x\right)$ is lower bounded by
\begin{equation}
\setlength\abovedisplayskip{5pt}
\setlength\belowdisplayskip{5pt}
\label{lemma1}
\ln \left( {1 + x} \right) \ge \rho \ln x + \delta ,
\end{equation}
where $\rho$ and $\delta$ are expressed as
\begin{equation}
\setlength\abovedisplayskip{5pt}
\setlength\belowdisplayskip{5pt}
\label{rho}
\rho  = \frac{{\hat x}}{{1 + \hat x}},\delta  = \ln \left( {1 + \hat x} \right) - \frac{{\hat x}}{{1 + \hat x}}\ln \left( {\hat x} \right).
\end{equation}

\emph{Proof}: Please refer to Appendix \ref{proof_lnx}. $\hfill\blacksquare$

\end{lemma}

\begin{lemma}
\label{vx}
 For any given ${\hat {x}} \ge \frac{{\sqrt {17}  - 3}}{4}$, function $G\left(x \right)$ always satisfies the following inequality:
\begin{equation}
\setlength\abovedisplayskip{5pt}
\setlength\belowdisplayskip{5pt}
\label{lemma2}
G\left( x \right) \le \tilde \rho \ln \left( x \right) + \tilde \delta ,
\end{equation}
where $\tilde \rho$ and $\tilde \delta $ are denoted by
\begin{equation}
\setlength\abovedisplayskip{5pt}
\setlength\belowdisplayskip{5pt}
\label{tilderho}
\tilde \rho  = \frac{{\hat x}}{{\sqrt {{{\hat x}^2} + 2\hat x} }} - \frac{{\hat x\sqrt {{{\hat x}^2} + 2\hat x} }}{{{{\left( {1 + \hat x} \right)}^2}}},
\end{equation}

and
\begin{equation}
\setlength\abovedisplayskip{5pt}
\setlength\belowdisplayskip{5pt}
\label{tildedelta}
\tilde \delta  = \sqrt {1 - \frac{1}{{{{\left( {1 + \hat x} \right)}^2}}}}  - \tilde \rho \ln \left( {\hat x} \right).
\end{equation}

\emph{Proof}: Please refer to Appendix D in \cite{ref13d}. $\hfill\blacksquare$

\end{lemma}

Based on Lemma \ref{lnx} and Lemma \ref{vx}, we can now solve the weighted sum rate maximization problem by using an iterative optimization algorithm. To this end, we first initialize the $k$th device's pilot power as $p_k^{p,{\left( 1 \right)}}$ and payload power as $p_k^{d,{\left( 1 \right)}}$, and calculate the corresponding SINR $\hat \gamma _k^{\rm{MRC}}$, which is denoted by ${\chi}_k^{\left( 1 \right)} = \hat \gamma _k^{\rm{MRC}}$. In the $i$th iteration, we approximate $\ln \left( 1+{\chi}_k \right)$ by $\ln \left( {{\chi _k}} \right) = {\rho ^{\left( i \right)}_{k}}\ln x + {\delta ^{\left( i \right)}_{k}}$ and $G \left( {{\chi _k}} \right) $ by $G \left( {{\chi _k}} \right) = {{\tilde \rho }^{\left( i \right)}_{k}}\ln \left( x \right) + {{\tilde \delta }^{\left( i \right)}_{k}}$, where ${\rho ^{\left( i \right)}_{k}}$ and ${\delta ^{\left( i \right)}_{k}}$ are obtained based on ({\ref{rho}) by using $\hat x={\chi}_k^{\left( i \right)}$, ${{\tilde \rho }^{\left( i \right)}_{k}}$ and ${{\tilde \delta }^{\left( i \right)}_{k}}$ are obtained based on ({\ref{tilderho}) and ({\ref{tildedelta}) by using $\hat x={\chi}_k^{\left( i \right)}$. As a result, the weighted sum rate can be lower bounded by
\begin{align}
\setlength\abovedisplayskip{5pt}
\setlength\belowdisplayskip{5pt}
& \sum\limits_{k = 1}^K {{w_k B}\frac{{\left( {1 - \eta } \right)}}{{\ln 2}}\left[ {\ln \left( {1 + {\chi _k}} \right) - {\alpha _k}G\left( {{\chi _k}} \right)} \right]} \notag \\
 \ge &\sum\limits_{k = 1}^K {{w_k B}\frac{{\left( {1 - \eta } \right)}}{{\ln 2}}\!\!\left[\! {{\rho ^{\left( i \right)}_{k}} \! \ln \left( {{\chi _k}} \right) \!+\! {\delta ^{\left( i \right)}_{k}} \!-\! {\alpha _k}{{\tilde \rho }^{\left( i \right)}_{k}}\ln \left( {{\chi _k}} \right) \!-\! {\alpha _k}{{\tilde \delta }^{\left( i \right)}_{k}}} \!\right]}  \notag \\
 = &\sum\limits_{k = 1}^K {{w_k B}\frac{{\left( {1 - \eta } \right)}}{{\ln 2}}\left[ {\ln {{\left( {{\chi _k}} \right)}^{\left[ {{\rho ^{\left( i \right)}_{k}} - {\alpha _k}{{\tilde \rho }^{\left( i \right)}_{k}}} \right]}} + {\delta ^{\left( i \right)}_{k}} - {\alpha _k}{{\tilde \delta }^{\left( i \right)}_{k}}} \right]} \label{equal_OF},
\end{align}
where the equality holds only when $\chi_k = \chi_k^{\left(i \right)}$.

Next, we optimize the LB of the objective function instead of the original objective function. Specifically, the subproblem to be solved in the $i$th iteration is given by
\begin{subequations}
\setlength\abovedisplayskip{5pt}
\setlength\belowdisplayskip{5pt}
\label{equal_formualtion}
\begin{align}
\mathop {\max }\limits_{\left\{ {p_k^p} \right\},\left\{ {p_k^d} \right\},\left\{ {{\chi _k}} \right\}} & \prod\limits_{k = 1}^K {{\chi _k}^{{{\hat w}_k^{\left( i \right)}}}}  \label{equal_formualtion_a}\\
{\text{s.t.}}\;\;\;\;\;\;\;\; & {\rm{(\ref{MRC_optimization_trans_b})}}, {\rm{(\ref{MRC_optimization_trans_c})}}, {\rm{(\ref{MRC_optimization_c})}} \label{equal_formualtion_b},
\end{align}
\end{subequations}
where ${{\hat w}^{\left( i \right)}_k}$ is equal to ${{\hat w}^{\left( i \right)}_k} = {w_k B}\frac{{\left( {1 - \eta } \right)}}{{\ln 2}}\left( {{\rho ^{\left( i \right)}} - {\alpha _k}{{\tilde \rho }^{\left( i \right)}}} \right)$. For the centralized mMIMO case in \cite{ref13d}, the above power allocation problem is a GP problem, which can be readily solved by using CVX tools. However, for the general case of user-centric CF mMIMO systems, the constraint (\ref{MRC_optimization_trans_b}) cannot be transformed into a GP form. As a result, the above problem is not a GP problem, which cannot be readily solved.

To address the abovementioned issue, we approximate the constraint (\ref{MRC_optimization_trans_b}) into a more tractable form by introducing the following lemma.
\begin{lemma}
\label{lemma_MRC_Sub}
The $k$th device's SINR $ \hat \gamma _k^{{\rm{MRC}}}$ can be rewritten as
\begin{equation}
\setlength\abovedisplayskip{5pt}
\setlength\belowdisplayskip{5pt}
\label{MRC_sinr_trans}
 \hat \gamma _k^{{\rm{MRC}}} = \frac {Np_{k}^d {\left( {{\theta _k}} \right)^2}} {{{\sigma _k}\left( {\sum\limits_{k' = 1}^K {{p^d_{k'}}} {\xi _{k,k'}} + {\theta _k}} \right)}},
\end{equation}
where ${\theta _k}$, ${\sigma _k}$, $\xi _{k,k'}$ are given by
\begin{equation}
\setlength\abovedisplayskip{5pt}
\setlength\belowdisplayskip{5pt}
\label{MRC_sinr_theta}
{\theta _k} = \sum\limits_{m \in {\mathcal{M}}_k} {\left[ {Kp_k^p{{\left( {{\beta _{m,k}}} \right)}^2}\prod\limits_{n \ne m} {\left( {Kp_k^p{\beta _{n,k}} + 1} \right)} } \right]},
\end{equation}
\begin{equation}
\setlength\abovedisplayskip{5pt}
\setlength\belowdisplayskip{5pt}
\label{MRC_sinr_sigma}
{\sigma _k} = \prod\limits_{m \in {\mathcal{M}}_k} {\left( {Kp_k^p{\beta _{m,k}} + 1} \right)},
\end{equation}
and
\begin{equation}
\setlength\abovedisplayskip{5pt}
\setlength\belowdisplayskip{5pt}
\label{MRC_sinr_xi}
{\xi _{k,k'}} = \sum\limits_{m \in {\mathcal{M}}_k} {\left[ {Kp_k^p{{\left( {{\beta _{m,k}}} \right)}^2}{\beta _{m,k'}}\prod\limits_{n \ne m} {\left( {Kp_k^p{\beta _{n,k}} + 1} \right)} } \right]}.
\end{equation}

\emph{Proof}: Please refer to Appendix \ref{proof_MRC_Substitution}. $\hfill\blacksquare$

\end{lemma}

By using Lemma \ref{lemma_MRC_Sub}, the constraint (\ref{MRC_optimization_trans_b}) can be reformulated as
\begin{equation}
\setlength\abovedisplayskip{5pt}
\setlength\belowdisplayskip{5pt}
\label{MRC_sinr_trans_appro}
Np_{k}^d {\left( {{\theta _k}} \right)^2} \ge \chi_k \left[ {{\sigma _k}\left( {\sum\limits_{k' = 1}^K {{p^d_{k'}}} {\xi _{k,k'}} + {\theta _k}} \right)} \right].
\end{equation}
However, both sides of (\ref{MRC_sinr_trans_appro}) are all posynomial functions, and thus constraint (\ref{MRC_sinr_trans_appro}) still does not satisfy the form of a GP problem. To deal with this difficulty, we utilize log-function to approximate $\theta_k$ into a monomial form as detailed in the following theorem.

\begin{theorem}
\label{theta_T}
For any given $\hat p_k^p > 0$, $\theta_k$ is lower bounded by
\begin{equation}
\setlength\abovedisplayskip{5pt}
\setlength\belowdisplayskip{5pt}
\label{MRC_theta_LB}
\begin{split}
	{\theta _k} &= Kp_k^p\sum\limits_{m \in {\mathcal{M}}_k} {\left[ {{{\left( {{\beta _{m,k}}} \right)}^2}\prod\limits_{n \ne m} {\left( {Kp_k^p{\beta _{n,k}} + 1} \right)} } \right]}  \\
	&\ge {{\rm{e}}^{{c_k}}}{\left( {p_k^p} \right)^{{a_k}}}  ,
\end{split}
\end{equation}
where ${\rm{e}}$ is the exponent, and $a_k$ and $c_k$ are given by
\begin{equation}
\setlength\abovedisplayskip{5pt}
\setlength\belowdisplayskip{5pt}
\label{a_k}
\begin{split}
{a_k} &= {\left. {\frac{{\partial \ln {\theta _k}}}{{\partial \ln \left( {p_k^p} \right)}}} \right|_{p_k^p = \hat p_k^p}} = 1 \\
  & +\! \frac{{\sum\limits_{m \in {\mathcal{M}}_k}\!\!\!{\left[ {{{\left( {{\beta _{m,k}}} \right)}^2}\!\!\sum\limits_{n \ne m}\!\! {\left[ {K{\beta _{n,k}}{\hat p^p_{k}}\!\!\prod\limits_{i \ne m,n}\!\! {\left( {K{\beta _{i,k}}{\hat p^p_{k}} + 1} \right)} } \right]} } \right]} }}{{\sum\limits_{m \in {\mathcal{M}}_k}\!\!\! {\left[ {{{\left( {{\beta _{m,k}}} \right)}^2}\prod\limits_{n \ne m}\!\! {\left( {K{\beta _{n,k}}{\hat p^p_{k}} + 1} \right)} } \right]} }} ,
\end{split}
\end{equation}
and
\begin{equation}
\setlength\abovedisplayskip{5pt}
\setlength\belowdisplayskip{5pt}
\label{c_k}
{c_k} = \ln \left( {{\hat \theta _k}} \right) - {a_k}\ln \left( {\hat p_k^p} \right),
\end{equation}
where $\hat \theta _k$ is obtained by substituting $p^p_k = \hat p^p_k$ into (\ref{MRC_sinr_theta}). Besides, it is obvious that the inequality in (\ref{MRC_theta_LB}) holds with equality only when $p^p_k = \hat p^p_k$.

\emph{Proof}: Please refer to Appendix \ref{proof_theta_T}. $\hfill\blacksquare$

\end{theorem}

Based on Theorem \ref{theta_T}, we replace the polynomial function $\theta _k$ in (\ref{MRC_theta_LB}) with the best local monomial approximations. Specifically, we use $a_k^{\left(i\right)}$ and $c_k^{\left(i\right)}$ to approximate $\theta _k$ in the $i$th iteration, then replace the left hand side of the inequality in (\ref{MRC_sinr_trans_appro}) by
\begin{equation}
\setlength\abovedisplayskip{5pt}
\setlength\belowdisplayskip{5pt}
\label{MRC_sinr_appro}
Np_{k}^d {\left( {{\theta _k}} \right)^2} \ge N{{ {{{ \rm {e}}^{2c_k^{\left( i \right)}}}{{\left( {{p^p_{k}}} \right)}^{2a_k^{\left( i \right)}}}} }}{p^d_k}.
\end{equation}

Through the above approximations, Problem (\ref{MRC_optimization}) for the MRC case is converted into a GP problem, which is given by
\begin{subequations}
\setlength\abovedisplayskip{5pt}
\setlength\belowdisplayskip{5pt}
\label{MRC_final}
\begin{align}
\mathop {\max }\limits_{\left\{ {p_k^p} \right\},\left\{ {p_k^d} \right\},\left\{ {{\chi _k}} \right\}} & \prod\limits_{k = 1}^K {{\chi _k}^{{{\hat w}_k^{\left( i \right)}}}}  \label{MRC_final_a} \\
{\text{s.t.}}\;\;\;\;\;\;\;\; & N{{ {{{ \rm {e}}^{2c_k^{\left( i \right)}}}{{\left( {{p^p_{k}}} \right)}^{2a_k^{\left( i \right)}}}} }}{p^d_k} \notag \\
&\ge {\chi _k} \left[ {{\sigma _k}\left( {\sum\limits_{k' = 1}^K {{p^d_{k'}}} {\xi _{k,k'}} + {\theta _k}} \right)} \right],\forall k , \label{MRC_final_b} \\
 & {\rm{(\ref{MRC_optimization_trans_c})}}, {\rm{(\ref{MRC_optimization_c})}} \label{MRC_final_c}.
\end{align}
\end{subequations}

For the iterative algorithm, we need to find a feasible solution to initialize the algorithm. To tackle this issue, we introduce an auxiliary variable $\varphi$ and construct an alternative optimization problem, which is given by
\begin{subequations}
\setlength\abovedisplayskip{5pt}
\setlength\belowdisplayskip{5pt}
\label{feasible_region}
\begin{align}
\mathop {\max } \limits_{\varphi ,\left\{ {p_k^p} \right\},\left\{ {p_k^d} \right\}}  & \varphi \label{feasible_region_a} \\
 {\text{s.t.}}\;\;\;\;\;  & N{{ {{{ \rm {e}}^{2c_k^{\left( i \right)}}}{{\left( {{p^p_{k}}} \right)}^{2a_k^{\left( i \right)}}}} }}{p^d_k} \notag \\
 & \ge  \frac{\varphi }{{f_k^{ - 1}\left( {\frac{{R_k^{{\rm{req}}}\ln 2}}{{(1 - \eta)B }}} \right)}} \left[ {{\sigma _k}\left( {\sum\limits_{k' = 1}^K {{p^d_{k'}}} {\xi _{k,k'}} + {\theta _k}} \right)} \right], \label{feasible_region_b} \\
& ({\rm{\ref{MRC_optimization_c}}}) \label{feasible_region_c}.
\end{align}
\end{subequations}
Obviously, Problem (\ref{feasible_region}) is always feasible. Similar to Problem (\ref{MRC_final}), Problem (\ref{feasible_region}) is also a GP problem, and the original Problem (\ref{MRC_final}) is feasible only if $\varphi$ is no smaller than 1. Based on the abovementioned discussion, the algorithm to solve Problem (\ref{MRC_final}) is given in Algorithm 1.
\begin{algorithm}[t]
\caption{Iterative Algorithm for Solving Problem (\ref{MRC_final}) for the MRC case}
\begin{algorithmic}[1]
\label{MRC_algorithm}
\STATE Initialize iteration number $i = 1$, and error tolerance $\zeta = 0.01$;
\STATE Initialize the pilot power and transmission power $ \left\{p^{p,\left( 1\right)}_k,p^{d,\left( 1\right)}_k,\forall k \right\}$ by solving Problem (\ref{feasible_region}), calculate SINR $\left \{\chi _k^{\left( 1 \right)},\forall k\right\}$, obtain the weighted sum rate in (\ref{MRC_optimization_a}) and denoted as ${\rm{Obj}}^{\left(1\right)}$. Set ${\rm{Obj}}^{\left(0\right)} = {\rm{Obj}}^{\left(1\right)} \zeta$;
\WHILE {${{\left( {{\rm{Ob}}{{\rm{j}}^{\left( i \right)}} - {\rm{Ob}}{{\rm{j}}^{\left( {i - 1} \right)}}} \right)} \mathord{\left/
 {\vphantom {{\left( {{\rm{Ob}}{{\rm{j}}^{\left( i \right)}} - {\rm{Ob}}{{\rm{j}}^{\left( {i - 1} \right)}}} \right)} {{\rm{Ob}}{{\rm{j}}^{\left( {i - 1} \right)}}}}} \right.
 \kern-\nulldelimiterspace} {{\rm{Ob}}{{\rm{j}}^{\left( {i - 1} \right)}}}} \ge \zeta$}
\STATE Update $\left \{ {{\hat w}^{\left( i \right)}_k},c^{\left(i\right)}_k, a^{\left(i\right)}_k,\forall k \right\}$;
\STATE Update $i = i+1$, solve Problem (\ref{MRC_final}) by using the CVX package to obtain $\left \{ p_k^{p,\left( i \right)},p_k^{d,\left( i \right)} ,\forall k\right\}$, calculate SINR $\left \{\chi _k^{\left( i \right)},\forall k\right\}$ and then obtain the weighted sum rate, denoted as ${\rm{Obj}}^{\left(i\right)}$;
\ENDWHILE
\end{algorithmic}
\end{algorithm}

\subsection{Algorithm Analysis}
1) \emph{Convergence Analysis}:
Before proving the convergence of our proposed algorithm, we first need to prove that the solution in the $i$th iteration is also feasible in the $(i+1)$th iteration. For Algorithm \ref{MRC_algorithm}, we only need to check whether constraint (\ref{MRC_final_b}) still holds since the constraints (\ref{MRC_optimization_trans_c}) and (\ref{MRC_optimization_c}) are the same in each iteration. The constraint (\ref{MRC_final_b}) in the $i$th iteration can be expressed as
\begin{equation}
\setlength\abovedisplayskip{5pt}
\setlength\belowdisplayskip{5pt}
\label{MRC_sinr_appro_n}
\begin{split}
	&N{{ {{{ \rm {e}}^{2c_k^{\left( i - 1\right)}}}{{\left( {{p^{p,\left( i \right)}_{k}}} \right)}^{2a_k^{\left( i - 1 \right)}}}} }}{p^{d,\left( i\right)}_k} \\
	\ge& {\chi _k^{\left( i\right)}} \left[ {{\sigma _k^{\left( i\right)}}\left( {\sum\limits_{k' = 1}^K {{p^{d,\left( i\right)}_{k'}}} {\xi ^{\left( i\right)}_{k,k'}} + {\theta _k^{\left( i\right)}}} \right)} \right],
\end{split}
\end{equation}
where $\left\{ \chi _k^{\left( i \right)}, p_{k}^{d,\left( i \right)}, p_{k}^{p,\left( i \right)}, \forall k\right\}$ is the optimal solution in the $i$th iteration.

By using Theorem \ref{theta_T} and ({\ref{MRC_sinr_appro}}), we have
\begin{equation}
\setlength\abovedisplayskip{5pt}
\setlength\belowdisplayskip{5pt}
\label{MRC_sinr_appro_n1}
\begin{split}
	&N{{ {{{ \rm {e}}^{2c_k^{\left( i\right)}}}{{\left( {{p^{p,\left( i \right)}_{k}}} \right)}^{2a_k^{\left( i\right)}}}} }}{p^{d,\left( i\right)}_k} = Np_{k}^{d,\left( i \right)} {\left( {{\theta _k^{\left( i \right)}}} \right)^2} \\
	\ge& N{{ {{{ \rm {e}}^{2c_k^{\left( i - 1\right)}}}{{\left( {{p^{p,\left( i \right)}_{k}}} \right)}^{2a_k^{\left( i - 1 \right)}}}} }}{p^{d,\left( i\right)}_k}.
\end{split}
\end{equation}
Then, by combining (\ref{MRC_sinr_appro_n}) with (\ref{MRC_sinr_appro_n1}), we have
\begin{equation}
\setlength\abovedisplayskip{5pt}
\setlength\belowdisplayskip{5pt}
\label{MRC_sinr_appro_n2}
\begin{split}
	&N{{ {{{ \rm {e}}^{2c_k^{\left( i\right)}}}{{\left( {{p^{p,\left( i \right)}_{k}}} \right)}^{2a_k^{\left( i\right)}}}} }}{p^{d,\left( i\right)}_k} \\
	\ge& {\chi _k^{\left( i\right)}} \left[ {{\sigma _k^{\left( i\right)}}\left( {\sum\limits_{k' = 1}^K {{p^{d,\left( i\right)}_{k'}}} {\xi ^{\left( i\right)}_{k,k'}} + {\theta _k^{\left( i\right)}}} \right)} \right].
\end{split}
\end{equation}
Therefore, we prove that the solution is also feasible for the solution in the $(i+1)$th iteration.

Finally, we denote ${\rm{Obj}}^{\left(i \right)}$ as the weighted sum rate in the $i$th iteration and prove the convergence of Algorithm \ref{MRC_algorithm}. Since the solution in the $i$th iteration is just a feasible solution in the $(i+1)$th iteration, we have
\begin{equation}
\setlength\abovedisplayskip{5pt}
\setlength\belowdisplayskip{5pt}
 \label{MRC_Cov1}
\begin{split}
 &\sum\limits_{k = 1}^K {{w_k B}\frac{{\left( {1 - \eta } \right)}}{{\ln 2}} \left[\! {\ln {{\left( {{\chi ^{\left( i + 1 \right)}_k}} \right)}^{\left[ {{\rho ^{\left( i \right)}_{k}} - {\alpha _k}{{\hat \rho }^{\left( i \right)}_{k}}} \right]}} + {\delta ^{\left( i \right)}_{k}} - {\alpha _k}{{\hat \delta }^{\left( i \right)}_{k}}} \!\right]} \\
\ge & \sum\limits_{k = 1}^K {{w_k B}\frac{{\left( {1 - \eta } \right)}}{{\ln 2}}  \left[\! {\ln {{\left( {{\chi ^{\left( i \right)}_k}} \right)}^{\left[ {{\rho ^{\left( i \right)}_{k}} - {\alpha _k}{{\hat \rho }^{\left( i \right)}_{k}}} \right]}} + {\delta ^{\left( i \right)}_{k}} - {\alpha _k}{{\hat \delta }^{\left( i \right)}_{k}}} \!\right]} \\
 =  & {\rm{Obj}}^{\left(i\right)},
 \end{split}
\end{equation}
where $\left\{\chi_k^{\left(i+1\right)},\forall k\right\}$ is the optimal solution to Problem (\ref{MRC_final}) in the $\left(i+1\right)$th iteration.

Substituting ${\chi _k }= {\chi ^{\left( i + 1 \right)}_k} $ into the inequality in (\ref{equal_OF}), we have
\begin{equation}
\setlength\abovedisplayskip{5pt}
\setlength\belowdisplayskip{5pt}
\label{MRC_inequality}
\begin{split}
& \sum\limits_{k = 1}^K {{w_k B}\frac{{\left( {1 - \eta } \right)}}{{\ln 2}}\left[ {\ln \left( {1 + {\chi ^{\left( i + 1 \right)}_k} } \right) - {\alpha _k}G\left( {{\chi ^{\left( i + 1 \right)}_k} } \right)} \right]}  \\
\ge  & \sum\limits_{k = 1}^K {{w_k B}\frac{{\left( {1 - \eta } \right)}}{{\ln 2}} \ln {{\left( {{\chi ^{\left( i + 1 \right)}_k}} \right)}^{\left[ {{\rho ^{\left( i + 1 \right)}_{k}} - {\alpha _k}{{\hat \rho }^{\left( i + 1\right)}_{k}}} \right]}}} \\
 &+  \sum\limits_{k = 1}^K{w_k B\frac{{\left( {1 - \eta } \right)}}{{\ln 2}} [{\delta ^{\left( i + 1 \right)}_{k}} - {\alpha _k}{{\hat \delta }^{\left( i + 1\right)}_{k}}] } \\
\ge &\sum\limits_{k = 1}^K {{w_k B}\frac{{\left( {1 - \eta } \right)}}{{\ln 2}}  \!\! \left[ \! {\ln {{\left( {{\chi ^{\left( i  \!+ \! 1 \right)}_k}} \right)}^{\left[ {{\rho ^{\left( i \right)}_{k}}  \!- \! {\alpha _k}{{\hat \rho }^{\left( i \right)}_{k}}} \right]}}  \!+ \! {\delta ^{\left( i \right)}_{k}}  \!- \! {\alpha _k}{{\hat \delta }^{\left( i \right)}_{k}}}  \!\right]} .
 \end{split}
\end{equation}

Then, the convergence of Algorithm \ref{MRC_algorithm} is verified by combining (\ref{MRC_Cov1}) with (\ref{MRC_inequality}), which can be expressed as
\begin{equation}
\setlength\abovedisplayskip{5pt}
\setlength\belowdisplayskip{5pt}
\label{MRC_Cov2}
\begin{split}
&{\rm{Obj}}^{\left(i+1\right)} \\
= & \sum\limits_{k = 1}^K {{w_k B}\frac{{\left( {1 - \eta } \right)}}{{\ln 2}} \left[\! {\ln \left( {1 + {\chi ^{\left( i + 1 \right)}_k}} \right) - {\alpha _k}G\left( {{\chi ^{\left( i + 1 \right)}_k}} \right)} \!\right]} \\
\ge & \sum\limits_{k = 1}^K {{w_k B}\frac{{\left( {1 - \eta } \right)}}{{\ln 2}}  \left[\! {\ln {{\left( {{\chi ^{\left( i + 1 \right)}_k}} \right)}^{\left[ {{\rho ^{\left( i \right)}_{k}} \!-\! {\alpha _k}{{\hat \rho }^{\left( i \right)}_{k}}} \right]}} \!+ \!{\delta ^{\left( i \right)}_{k}} \!-\! {\alpha _k}{{\hat \delta }^{\left( i \right)}_{k}}} \!\right]} \\
\ge &  {\rm{Obj}}^{\left(i\right)}  .
 \end{split}
\end{equation}

Even though it is difficult to obtain the optimal solution of the non-convex Problem (\ref{MRC_optimization}), we can prove that Algorithm \ref{MRC_algorithm} can converge to the Karush-Kuhn-Tucker (KKT) point of Problem (\ref{MRC_optimization}) for the MRC decoder by using the similar proof as that in Appendix B in \cite{ref21}.

2) \emph{Complexity Analysis}: The complexity of Algorithm \ref{MRC_algorithm} mainly depends on the complexity of each iteration and the number of iterations. For the complexity of each iteration, the authors \textcolor{black}{of} \cite{ref22} claimed that the GP problem can be efficiently solved by using the standard interior point methods with a worst-case polynomial-time complexity. \textcolor{black}{Specifically, the main complexity of each iteration in Algorithm \ref{MRC_algorithm} lies in solving Problem (\ref{MRC_final}) which includes $3K$ variables and $3K$ constraints. Based on \cite{ref22}, the computational complexity of this algorithm is on the order of ${\mathcal{O}}(N_{iter} \times \max\{(3K)^{3}), N_{cost}\})$, where $N_{iter}$ is the number of iterations and $N_{cost}$ is the computational complexity of calculating the first-order and second-order derivatives of the objective function and constraint functions of Problem (\ref{MRC_final}) \cite{van2018joint}.} More importantly, simulation results show that Algorithm 1 converges rapidly, which demonstrates that Algorithm 1 can obtain a locally optimal solution with a polynomial time complexity.

\section{Power Allocation for the Case of FZF}
In this section, we aim to solve the weighted sum rate maximization problem for the case of the FZF decoder.

\subsection{Joint Optimization}
Different from the MRC decoder, the expression of SINR at \textcolor{black}{the} $k$th device by using FZF decoder is much more complicated. Before solving the optimization problem, we first rewrite the SINR's expression in a more tractable form as in the following lemma.

\begin{lemma}
\label{lemma_FZF_Substitution}
The $\hat \gamma _k^{{\rm{FZF}}}$ can be equivalently reformulated as
\begin{equation}
\setlength\abovedisplayskip{5pt}
\setlength\belowdisplayskip{5pt}
\label{ZF_SINR_equal}
\hat \gamma _k^{{\rm{FZF}}} = \frac{{{p^d_k}\left( {N - K} \right){{\left( {{\varpi _k}} \right)}^2}\prod\limits_{k' \ne k}^K {{{\left( {{\vartheta _{k,k'}}} \right)}^2}} }}{{|{\cal M}_k| {\prod\limits_{k' = 1}^K} {{{\left( {{\vartheta _{k,k'}}} \right)}^2}}  + \sum\limits_{k' = 1}^K {\left[ {{p^d_{k'}}{\mu _{k,k'}}\prod\limits_{j \ne k'}^K {{{\left( {{\vartheta _{k,j}}} \right)}^2}} } \right]} }},
\end{equation}
where ${\varpi _k}$, $\vartheta _{k'}$, and ${\mu _{k'}}$ are given by
\begin{equation}
\setlength\abovedisplayskip{5pt}
\setlength\belowdisplayskip{5pt}
\label{ZF_varpi}
{\varpi _k} = {\sum\limits_{m \in {\mathcal{M}}_k} {\left[ {\sqrt {K{p^p_{k}}{{\left( {{\beta _{m,k}}} \right)}^2}} \prod\limits_{n \ne m} {\sqrt {K{p^p_{k}}{\beta _{n,k}} + 1} } } \right]} },
\end{equation}

\begin{equation}
\setlength\abovedisplayskip{5pt}
\setlength\belowdisplayskip{5pt}
\label{ZF_vartheta}
\vartheta _{k,k'} = {\prod\limits_{m \in {\mathcal{M}}_k} {\sqrt {K{p^p_{k'}}{\beta _{m,k'}} + 1} } },
\end{equation}
and
\begin{equation}
\setlength\abovedisplayskip{5pt}
\setlength\belowdisplayskip{5pt}
\label{ZF_mu}
{\mu _{k,k'}} = {\sum\limits_{m \in {\mathcal{M}}_k} {\left[ {{\beta _{m,k'}}\prod\limits_{n \ne m} {\left( {K{p^p_{k'}}{\beta _{n,k'}} + 1} \right)} } \right]} }.
\end{equation}

\emph{Proof}: Please refer to Appendix \ref{proof_FZF_Substitution}. $\hfill\blacksquare$

\end{lemma}



It is readily found that the numerator in (\ref{ZF_SINR_equal}) is not a monomial function, and thus Problem (\ref{MRC_optimization}) for the FZF case is not a GP problem. Note that the numerator of the SINR at \textcolor{black}{the} $k$th device by using the FZF decoder is much more complicated than the case of the MRC decoder, and the approximate method for the MRC decoder cannot be directly applied to the FZF decoder. To address this issue, we introduce the following theorem.

\begin{theorem}
\label{varpi_T}
For any given pilot power ${\hat { \bf {P}}}^p = [{\hat p}^p_1,{\hat p}^p_2, \cdot \cdot \cdot, {\hat p}^p_K]$ with ${\hat p}^p_k > 0$, ${{\left( {{\varpi _k}} \right)}^2}\prod\limits_{k' \ne k}^K {{{\left( {{\vartheta _{k,k'}}} \right)}^2}} $ is lower bounded by
\begin{equation}
\setlength\abovedisplayskip{5pt}
\setlength\belowdisplayskip{5pt}
\label{ZF_inequality}
{{\left( {{\varpi _k}} \right)}^2}\prod\limits_{k' \ne k}^K {{{\left( {{\vartheta _{k,k'}}} \right)}^2}} \ge {{\rm e} ^{d_k}} \prod\limits_{j = 1}^K {{{\left( {p_j^p} \right)}^{{b^k_j}}}},
\end{equation}
where ${d_k}$ and ${{b^k_j}}$ are given by
\begin{equation}
\setlength\abovedisplayskip{5pt}
\setlength\belowdisplayskip{5pt}
\label{ZF_bk}
{b^k_j} = \left\{ {\begin{array}{*{20}{c}}
{{\left. {2\frac{{\partial \ln \left( {{\vartheta _{k,j}}} \right)}}{{\partial \ln \left( {p_j^p} \right)}}} \right|_{p_j^p = \hat p_j^p}},}&{j \ne k}\\
{2{\left. {\frac{{\partial \ln \left( {{\varpi _k}} \right)}}{{\partial \ln \left( {p_k^p} \right)}}} \right|_{p_k^p = \hat p_k^p}},}&{j = k}
\end{array}} \right.,
\end{equation}
and
\begin{equation}
\setlength\abovedisplayskip{5pt}
\setlength\belowdisplayskip{5pt}
\label{ZF_Ck}
{d_k} = \ln \left( {{{\left( {{\hat \varpi _k}} \right)}^2}\prod\limits_{k' \ne k}^K {{{\left( {{\hat \vartheta _{k,k'}}} \right)}^2}} } \right) - \sum\limits_{j = 1}^K {{b^k_j}\ln \left( {\hat p_j^p} \right)},
\end{equation}
where  $\hat \varpi _k$ and  $\hat \vartheta _{k,k'}$ are obtained by substituting $p_k^p =\hat p_k^p$ into (\ref{ZF_varpi}) and (\ref{ZF_vartheta}), respectively. In addition, the inequality in (\ref{ZF_inequality}) holds with equality only when $p_k^p=\hat p_k^p$.

\emph{Proof}: Please refer to Appendix \ref{proof_varpi_T}. $\hfill\blacksquare$

\end{theorem}

According to Theorem \ref{varpi_T}, the posynomial ${{\left( {{\varpi _k}} \right)}^2}\prod\limits_{k' \ne k}^K {{{\left( {{\vartheta _{k,k'}}} \right)}^2}} $ can be replaced by the best local monomial approximation. Specifically, the updated $d_k^{\left(i\right)}$ and $ {{\bf{b}}^{k,\left( i \right)}} = {\rm{ }}\left[ {{b^{k,\left( i \right)}_1},{b^{k,\left( i \right)}_2}, \cdot  \cdot  \cdot ,{b^{k,\left( i \right)}_K}} \right]$ are utilized to approximate the numerator of $\hat \gamma _k^{\rm{FZF}}$ in the $i$th iteration, denoted as
\begin{equation}
\setlength\abovedisplayskip{5pt}
\setlength\belowdisplayskip{5pt}
\label{appro_FZF_b}
\begin{split}
	&{{p^d_k}\left( {N - K} \right){{\left( {{\varpi _k}} \right)}^2}\prod\limits_{k' \ne k}^K {{{\left( {{\vartheta _{k,k'}}} \right)}^2}} }  \\ 
	\ge &{p^d_k}\left( {N - K} \right) {{\rm e} ^{d_k^{\left(i\right)}}} \prod\limits_{j = 1}^K {{{\left( {p_j^p} \right)}^{{b_j^{k,\left(i\right)}}}}}.
\end{split}
\end{equation}

Then, the original SINR constraint in (\ref{MRC_optimization_trans_b}) can be replaced by the following constraint
\begin{equation}
\setlength\abovedisplayskip{5pt}
\setlength\belowdisplayskip{5pt}
\label{FZF_trans}
\begin{split}
	&{p^d_k}\left( {N - K} \right) {{\rm e} ^{d_k^{\left(i\right)}}} \prod\limits_{j = 1}^K {{{\left( {p_j^p} \right)}^{{b_j^{k,\left(i\right)}}}}} \\
	\ge &{\chi _k} \left\{ \!\!{\sum\limits_{m \in {\mathcal{M}}_k}{\prod\limits_{k' = 1}^K} {{{\left( {{\vartheta _{k,k'}}} \right)}^2}}  + \sum\limits_{k' = 1}^K {\left[ {{p^d_{k'}}{\mu _{k,k'}}\prod\limits_{j \ne k'}^K {{{\left( {{\vartheta _{k,j}}} \right)}^2}} } \right]} }\!\! \right \}.
\end{split}
\end{equation}

Based on the above analysis, the optimization problem can be transformed into the following GP problem
\begin{subequations}
\setlength\abovedisplayskip{5pt}
\setlength\belowdisplayskip{5pt}
\label{ZF_optimization}
\begin{align}
\mathop {\max }\limits_{\left\{ {p_k^p} \right\},\left\{ {p_k^d} \right\},\left\{ {{\chi _k}} \right\}} & \prod\limits_{k = 1}^K {{\chi _k}^{{{\hat w}_k^{\left( i \right)}}}}   \label{ZF_optimization_a}\\
{\rm{s}}{\rm{.t}}{\rm{.}} & \left(\ref{FZF_trans}\right),\forall k,  \label{ZF_optimization_b} \\
& {\rm{(\ref{MRC_optimization_trans_c})}}, {\rm{(\ref{MRC_optimization_c})}}. \label{ZF_optimization_c}
\end{align}
\end{subequations}
Therefore, an iterative algorithm is proposed to solve Problem (\ref{ZF_optimization}), which is shown in Algorithm 2. In addition, the initialization scheme similar to the MRC case can be adopted to find a feasible initial point for Algorithm \ref{algo_FZF}. The convergence of Algorithm \ref{algo_FZF} can be readily proved by using the similar method for the MRC decoder, which is omitted due to limited space.

\begin{algorithm}[t]
\caption{Iterative Algorithm for Solving Problem (\ref{ZF_optimization}) for the FZF case}
\begin{algorithmic}[1]
\label{algo_FZF}
\STATE Initialize iteration number $i = 1$, and error tolerance $\zeta = 0.01$;
\STATE Initialize the pilot power and transmission power $ \left\{p^{p,\left( 1\right)}_k,p^{d,\left( 1\right)}_k,\forall k \right\}$, calculate SINR $\left \{ {\chi}_k^{\left( 1 \right)}, \forall k\right\}$, and obtain the weighted sum rate that is ${\rm{Obj}}^{\left(1\right)}$. Set ${\rm{Obj}}^{\left(0\right)} = {\rm{Obj}}^{\left(1\right)} \zeta$;
\WHILE {${{\left( {{\rm{Ob}}{{\rm{j}}^{\left( i \right)}} - {\rm{Ob}}{{\rm{j}}^{\left( {i - 1} \right)}}} \right)} \mathord{\left/
 {\vphantom {{\left( {{\rm{Ob}}{{\rm{j}}^{\left( i \right)}} - {\rm{Ob}}{{\rm{j}}^{\left( {i - 1} \right)}}} \right)} {{\rm{Ob}}{{\rm{j}}^{\left( {i - 1} \right)}}}}} \right.
 \kern-\nulldelimiterspace} {{\rm{Ob}}{{\rm{j}}^{\left( {i - 1} \right)}}}} \ge \zeta$}
\STATE Update $\left \{ {{\hat w}^{\left( i \right)}_k},d^{\left(i\right)}_k, {\bf{b}}^{k,\left(i\right)},\forall k \right\}$;
\STATE Update $i = i+1$, solve Problem (\ref{ZF_optimization}) by using the CVX package to obtain $\left \{ p_k^{p,\left( i \right)},p_k^{d,\left( i \right)} ,\forall k\right\}$, calculate SINR $\left \{\chi _k^{\left( i \right)},\forall k\right\}$ and then obtain the weighted sum rate, denoted as ${\rm{Obj}}^{\left(i\right)}$;
\ENDWHILE
\end{algorithmic}
\end{algorithm}

\section{Simulation Results}
In this section, we provide simulation results to demonstrate the effectiveness of our proposed algorithms for a smart factory. The factory is assumed to be a square with size of 1 km  $\times$ 1 km, and all the APs are uniformly deployed at $M$ constellation points. For the large-scale fading, we adopt three slope model for path-loss $\left(\rm{dB}\right)$ \cite{ref16}, which is given by
\begin{equation}
	\setlength\abovedisplayskip{5pt}
	\setlength\belowdisplayskip{5pt}
	\label{channel_model}
	{\rm{P}}{{\rm{L}}_{m,k}} = \left\{ {\begin{array}{*{20}{l}}
			{\begin{array}{*{20}{l}}
					{L_{\rm{loss}} \!+\! 35{{\log }_{10}}\left( {{d_{m,k}}} \right),{{d_{m,k}} \!>\! {d_1}}, }\\
					{L_{\rm{loss}} \!+\! 15{{\log }_{10}}\left( {{d_1}} \right) \!+\! 20{{\log }_{10}}\left( {{d_0}} \right),{{d_{m,k}} \!\le\! {d_0}},} \\
					{L_{\rm{loss}} \!+\! 15{{\log }_{10}}\left( {{d_1}} \right) \!+\! 20{{\log }_{10}}\left( {{d_{m,k}}} \right), \rm{other},}\\
			\end{array}}
	\end{array}} \right.
\end{equation}
where $d_{m,k}$ is the distance between the $m$th AP and the $k$th device, and $L_{\rm{loss}}$ is a constant factor, denoted as
\begin{equation}
\setlength\abovedisplayskip{5pt}
\setlength\belowdisplayskip{5pt}
\label{path_loss_L}
\begin{split}
L_{\rm{loss}} &= 46.3 + 33.9{\log _{10}}\left( f \right) - 13.82{\log _{10}}\left( {{h_{{\rm{AP}}}}} \right) \\
 & - \left( {1.1{{\log }_{10}}\left( f \right) - 0.7} \right){h_u} + \left( {1.56{{\log }_{10}}\left( f \right) - 0.8} \right) ,
\end{split}
\end{equation}
where $f\left(\rm{MHz}\right)$ is the carrier frequency, and $h_{{\rm{AP}}} \left(\rm{m}\right)$ and $h_u \left(\rm{m}\right)$ are the heights of the APs and devices, respectively. For the small-scale fading, it is generally modeled as Rayleigh fading with zero mean and unit variance. Unless otherwise specified, the simulation parameters are summarized in Table I. The noise power is given by
\begin{equation}
\setlength\abovedisplayskip{5pt}
\setlength\belowdisplayskip{5pt}
\label{noise_power}
P_n = B \times {k_B} \times {T_0} \times  10^{\frac{N_{\rm{dB}}}{10}} \left( {\rm{W}} \right),
\end{equation}
where $k_B = 1.381 \times 10^{-23}$ (Joule per Kelvin) is the Boltzmann constant, and ${T_0} = 290$ (Kelvin) is the noise temperature. The weights for all the devices are randomly generated within [0,1]. More importantly, we assume that the total number of antennas in this square area is constant. In other words, if there are more APs, each AP is equipped with less antennas.

Due to the implementation complexity, each device cannot be served by all APs, and thus several nearest APs are chosen to provide the service for each device. Inspired by \cite{ref14}, the following strategy is adopted
\begin{equation}
\setlength\abovedisplayskip{5pt}
\setlength\belowdisplayskip{5pt}
\label{AP_selection}
\frac{\sum \nolimits_{m \in {\mathcal{M}}_k} {\beta_{m,k}}}{\sum\nolimits_{m = 1}^M {{\beta _{m,k}}}} \ge  T_h,
\end{equation}
where $T_h$ is the threshold. Specifically, the large-scale fading parameters are arranged in descending order, and then select them in turn until the above condition in (\ref{AP_selection}) is satisfied. For example, $T_h = 1$ means each device is served by all APs.

\begin{table*}[t]
        \small
        \caption{Simulation Parameters}
		\centering
		\begin{tabular}{|c|c|c|c|}\hline
			Parameters Setting & Value & Parameters Setting & Value \\ \hline
            Carrier frequency ($f$) & 2.1 GHz & Bandwidth ($B$) & 10 MHz \\
            Transmission duration ($T_B$) & 0.1 ms & Blocklength ($L = B T_B$) & 1000 \\
            Height of APs (${h_{{\rm{AP}}}}$)& 15 m & Height of devices ($h_u$) & 1.6 m \\
			Noise figure ($N_{\rm{dB}}$) & 9 dB &  Number of devices ($K$) & 10 \\
            Required data rate ($R_{\rm{req}}$) & 5 Mbps & Decoding error probability ${{\varepsilon _k}}$ & $10^{-5}$ \\
            $d_{0}$ & 10 m & $d_{1}$ & 50 m \\ \hline
		\end{tabular}
		\label{tab: Margin_settings}
\end{table*}

\subsection{Tightness of the Date Rate LB}
\textcolor{black}{We first evaluate the gap between the derived LB and ergodic data rate for both MRC and FZF decoders with $T_h = 0.9$ and $p_k^d = p_k^p = 0.1$ $W$, $\forall k$. Simulation results are obtained through the Monte-Carlo simulation by averaging over $10^4$ random channel generations. As can be seen in Fig. \ref{fig:MRC_LB}, there is a gap between the derived LB and the ergodic rate for the MRC case, and the gap enlarges with the number of AP. Nevertheless, the analytical results have the same trend with the Monte Carlo simulations. In Fig. \ref{fig:FZF_LB}, we find that the derived LB based on the FZF decoder is close to the ergodic rate for any cases. This is due to the fact that the devices' interference is suppressed by the FZF decoder. More importantly, simulation results demonstrate that we can optimize the derived LB data rate instead of the complicated expectation expression.} 

\begin{figure*}[t]
\begin{minipage}[t]{0.49\linewidth}
\centering
\includegraphics[width=3.2in]{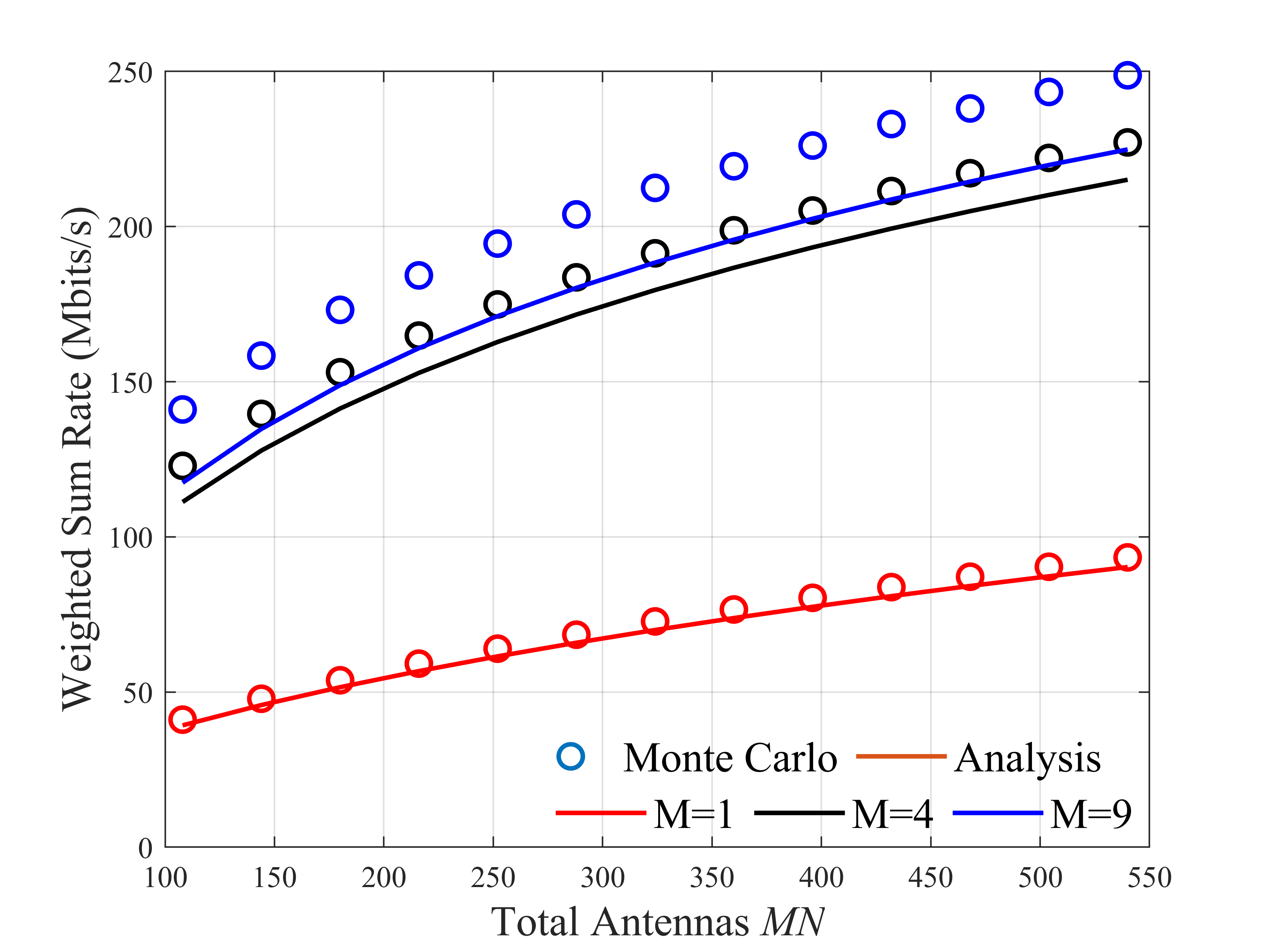}
\vspace{-0.2cm}
\caption{Weighted Sum Rate V.S. The Number of Total Antennas under various numbers of APs for the MRC decoder.}
\label{fig:MRC_LB}
\end{minipage}%
\hfill
\begin{minipage}[t]{0.49\linewidth}
\centering
\includegraphics[width=3.2in]{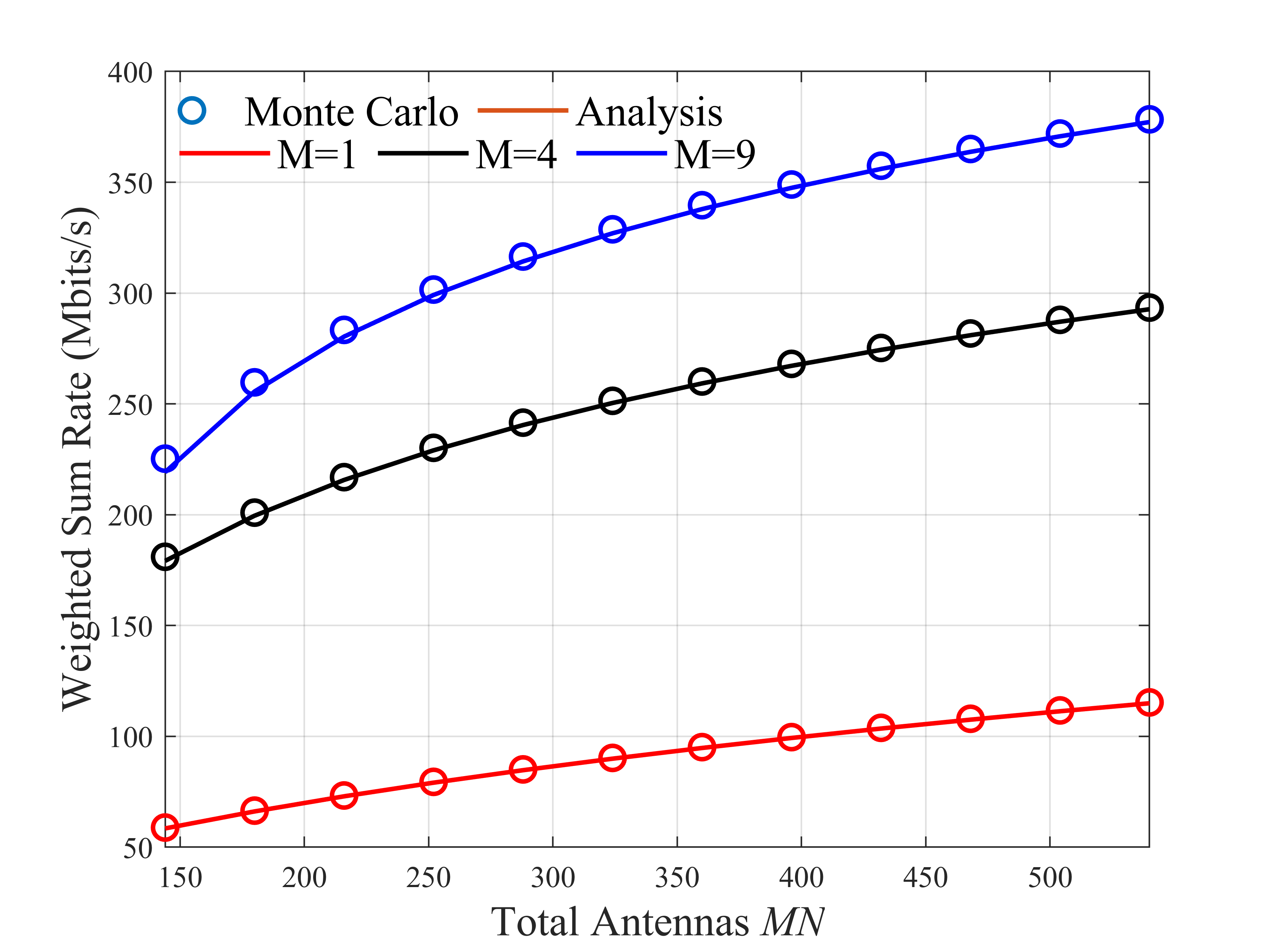}
\vspace{-0.2cm}
\caption{Weighted Sum Rate V.S. The Number of Total Antennas under various numbers of APs for the FZF decoder.}
\label{fig:FZF_LB}
\end{minipage}%
\end{figure*}

\subsection{Convergence of the Proposed Algorithms}
In this subsection, the devices are randomly deployed in the network, and then demonstrate the convergence of the proposed algorithms with a total number of antennas equal to $MN = 144$, illustrated in Fig. \ref{fig:MRC_converge} and Fig. \ref{fig:FZF_converge}. From these two figures, it is obvious that both algorithms converge rapidly regardless of the number of APs. Specifically, only 3 or 4 iterations are sufficient for the algorithm to converge for the MRC decoder, while the FZF decoder needs about 12 iterations to converge, which demonstrates the low complexity of our proposed algorithms.

\begin{figure*}
\begin{minipage}[t]{0.49\linewidth}
\centering
\includegraphics[width=3.2in]{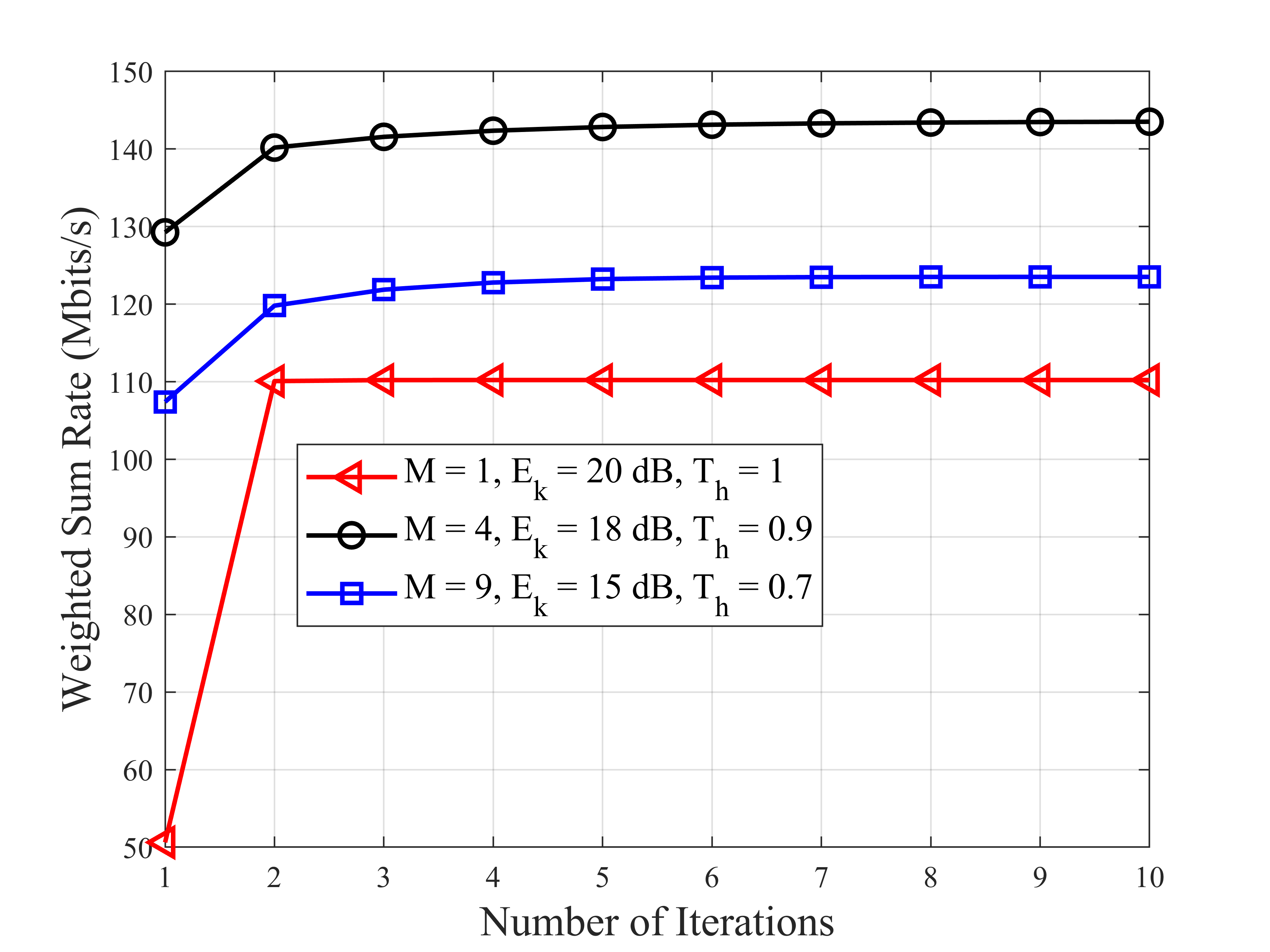}
\vspace{-0.2cm}
\caption{Convergence of the proposed algorithms for the MRC decoder.}
\label{fig:MRC_converge}
\end{minipage}%
\hfill
\begin{minipage}[t]{0.49\linewidth}
\centering
\includegraphics[width=3.2in]{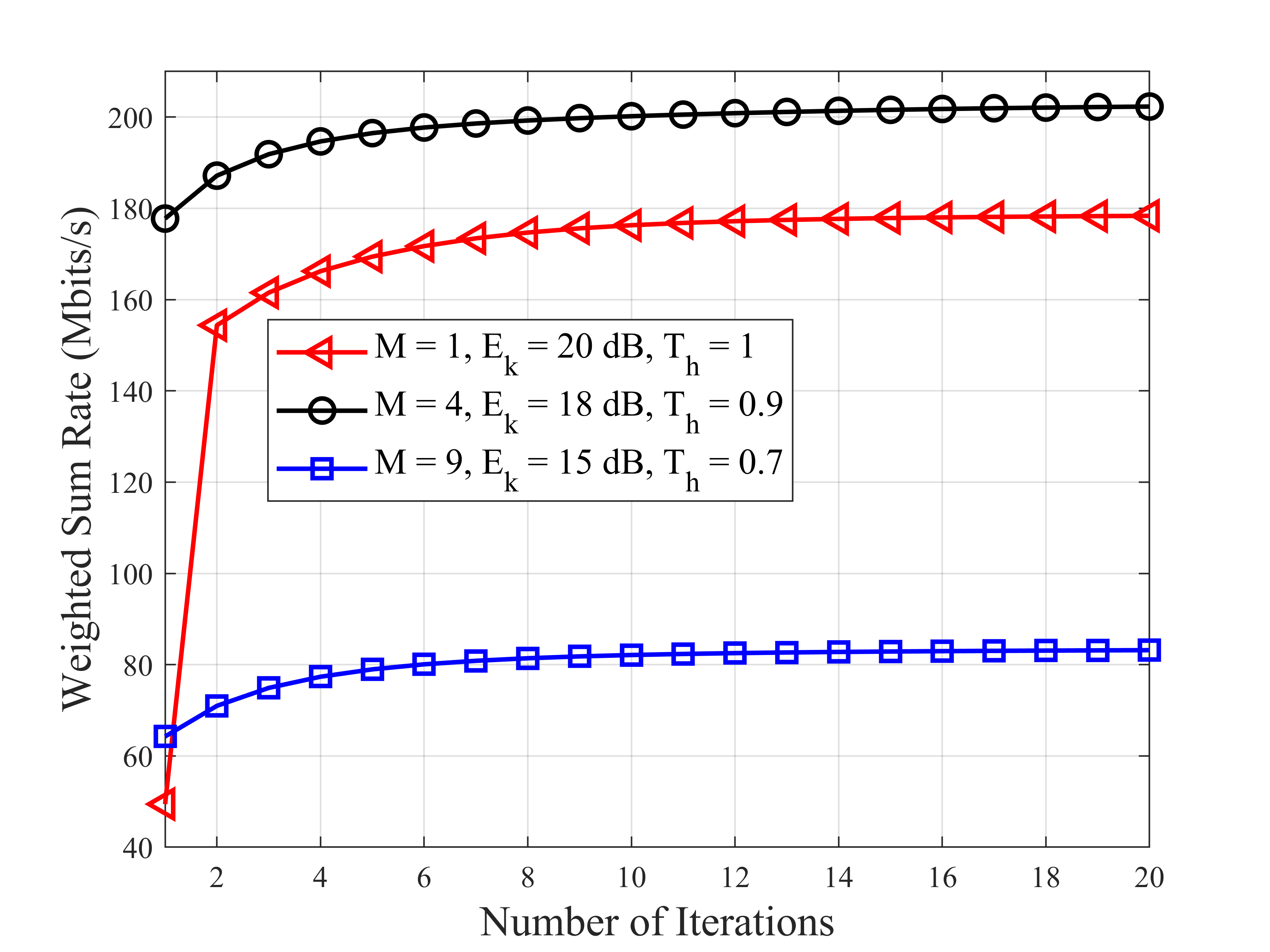}
\vspace{-0.2cm}
\caption{Converge of the proposed algorithms for the FZF decoder.}
\label{fig:FZF_converge}
\end{minipage}%
\end{figure*}

\subsection{Effect of the Threshold}
In this subsection, we demonstrate the impact of \textcolor{black}{the} threshold on the system performance for the AP selection. Besides, the average weighted sum rate is obtained by averaging over $10^2$ random deployments in the square area, and the system performance is defined as zero if any devices do not satisfy the strict requirements of data rate and DEP.

For the MRC case, the performance with different thresholds $T_h$ is illustrated in Fig. \ref{fig:MRC_VS_Threshold}. Initially, more APs can provide improved performance. However, the system performance deteriorates when choosing all APs, and the reason is that the APs that are far away from the device introduce unexpected interference. Besides, we find that the average performance can reach the peak performance when the threshold is about 0.95 for both 4 and 9 APs.

The impact of the threshold on the system performance for the FZF decoder is illustrated in Fig. \ref{fig:FZF_VS_Threshold}. Similar to the MRC decoder, the average weighted sum rate first increases with the threshold and then decreases with it. It can be seen from Fig. \ref{fig:FZF_VS_Threshold} that the optimal threshold is about $0.8$, and the performance would \textcolor{black}{deteriorate} after this threshold. Therefore, we set the threshold for the MRC and FZF decoders as 0.95 and 0.75 to reduce the implementation complexity.

\begin{figure*}
\begin{minipage}[t]{0.49\linewidth}
\centering
\includegraphics[width=3.2in]{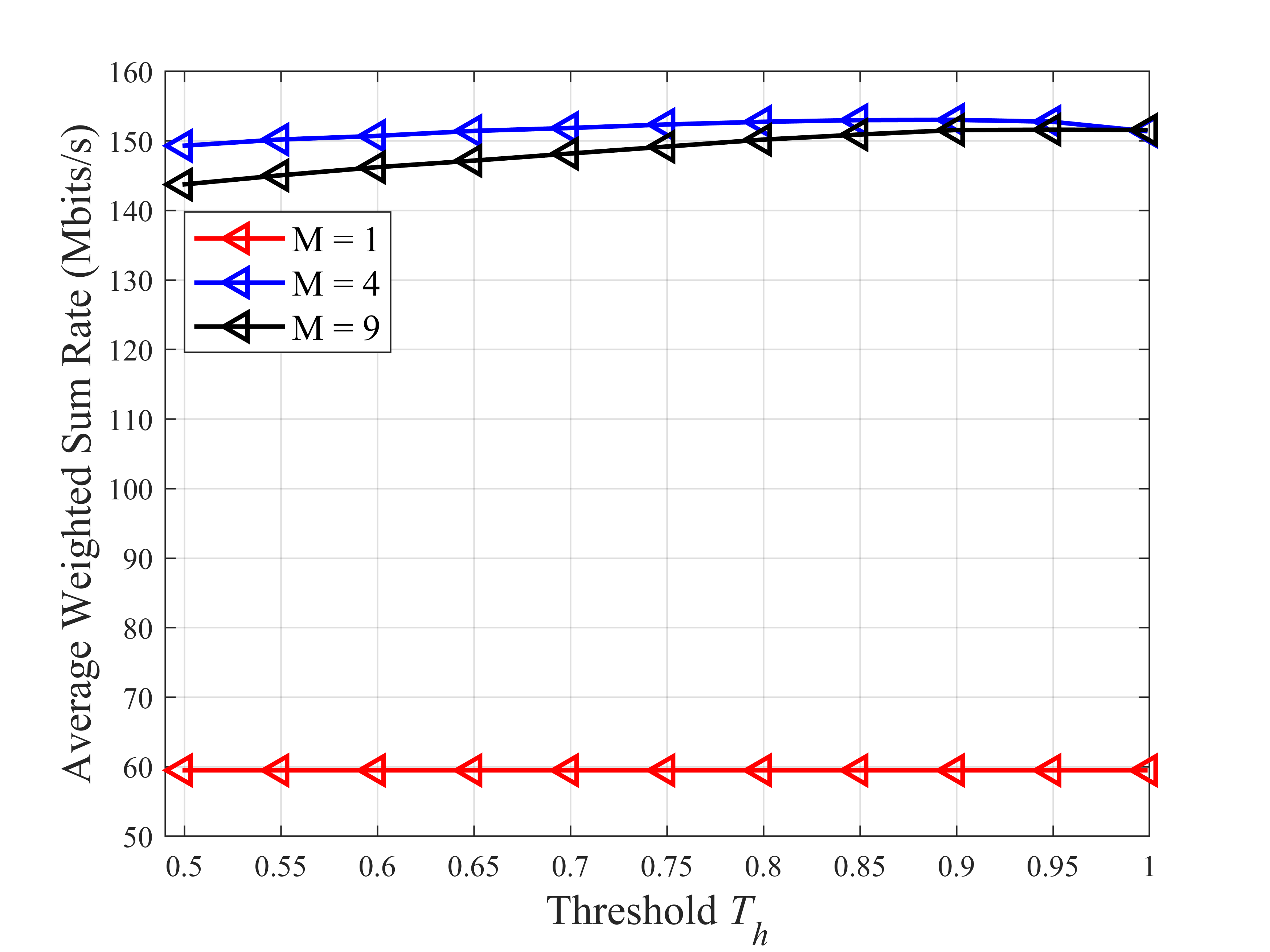}
\vspace{-0.2cm}
\caption{Average Weighted Sum Rate V.S. Threshold for MRC receiver.}
\label{fig:MRC_VS_Threshold}
\end{minipage}%
\hfill
\begin{minipage}[t]{0.49\linewidth}
\centering
\includegraphics[width=3.2in]{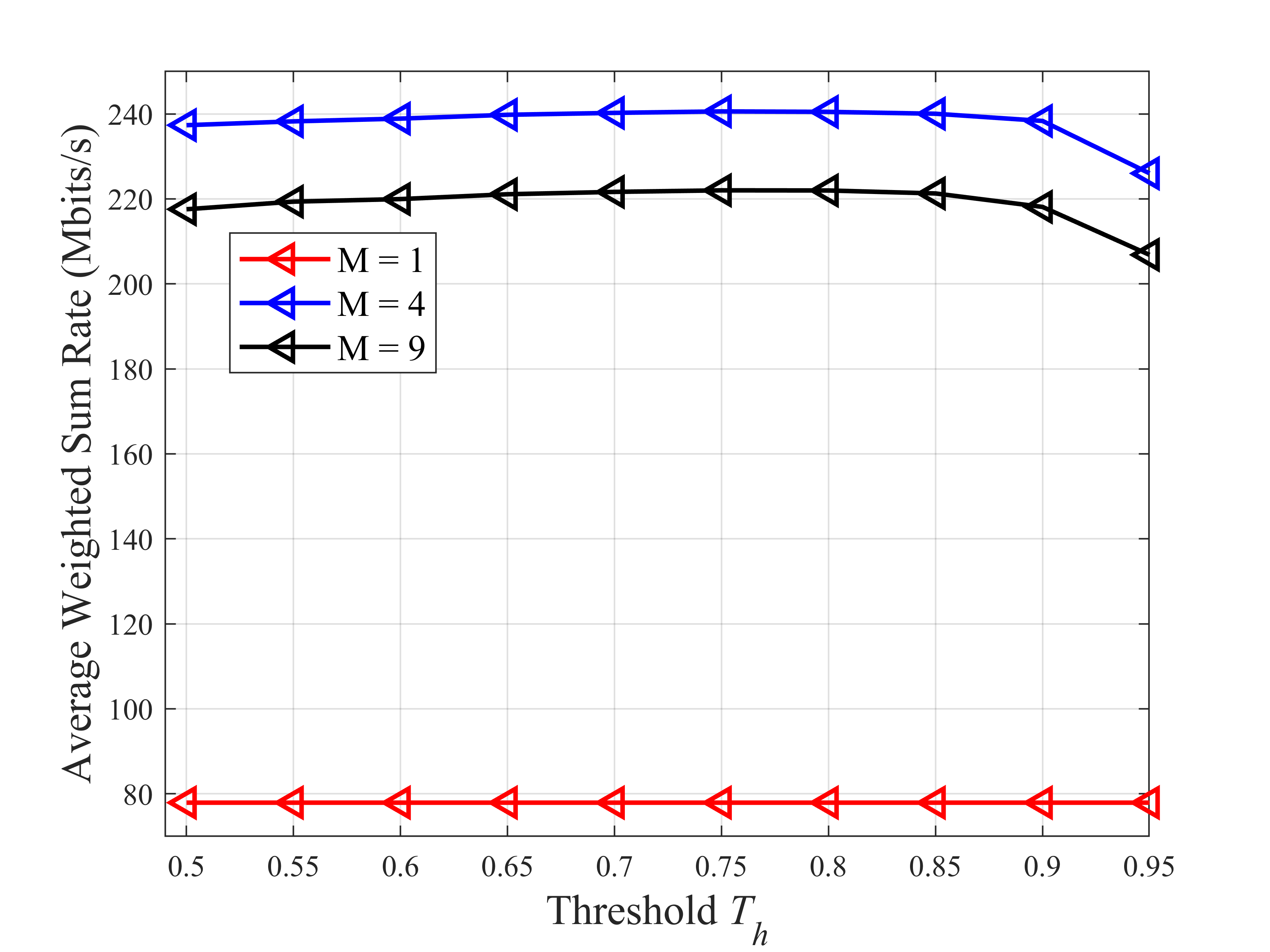}
\vspace{-0.2cm}
\caption{Average Weighted Sum Rate V.S. Threshold for FZF receiver.}
\label{fig:FZF_VS_Threshold}
\end{minipage}%
\end{figure*}

\subsection{Performance Comparison}
In this subsection, we investigate the impact of limited energy $E_k$ on the system performance. The total number of antennas $MN$ is set to $144$, and the simulation results are obtained by averaging over 100 Monte-Carlo trials where in each snapshot the devices' locations are randomly generated in the square area. Besides, to guarantee the URLLC requirements, we set the weight sum rate to zero when the achievable data rate of at least one device does not meet the required data rate. We also compare the proposed algorithms with the following algorithms.
\begin{itemize}
  \item {\bf{Upper bound}}: Shannon capacity \textcolor{black}{under infinite blocklength (IFBL)} is adopted for solving the problems in (\ref{MRC_optimization}) and (\ref{ZF_optimization}). Since the performance loss due to FBL is ignored, the Shannon capacity without penalty can be regarded as the upper bound of the average weighted sum rate in the smart factory.
  \item {\bf{Conventional alg.}}: We first obtain the solution based on \textcolor{black}{the} Shannon capacity \textcolor{black}{under IFBL}, and then the solution is applied to both MRC and FZF cases under \textcolor{black}{channel} capacity \textcolor{black}{in the FBL regime}, respectively. Specifically, we first obtain the optimal pilot power $p_k^p$ and payload power $p_k^d$ by solving the \textcolor{black}{conventional} Shannon capacity maximization problem, and then calculate the achievable data rate based on the FBL.
  \item {\bf{Fixed pilot power alg.}}: The pilot power is fixed as $p_k^p= \frac{E_k}{L}$ and we only optimize the payload power $p_k^d$ for maximizing the weighted sum rate.
\end{itemize}

In Fig. \ref{fig_MRC_performance_comparison}, we show the average weighted sum rate versus the total energy $E_k$ for the MRC decoder. It is obvious that the upper bound can obtain the best performance as the penalty due to FBL is ignored. Specifically, our proposed method is superior over the fixed pilot power algorithm, especially for the low energy regime. This is due to the fact the system performance is limited by \textcolor{black}{the} estimated the channel gain $\lambda_{m,k}$ in \textcolor{black}{the} low energy regime, and the proposed algorithm can flexibly allocate energy by jointly considering channel gain and transmission gain to maximize the system performance. In addition, the conventional method performs much worse than our proposed algorithm, because the optimization based on the \textcolor{black}{conventional} Shannon capacity does not consider the penalty due to FBL.

\begin{figure*}
\begin{minipage}[t]{0.32\linewidth}
{\includegraphics[width=2.25in]{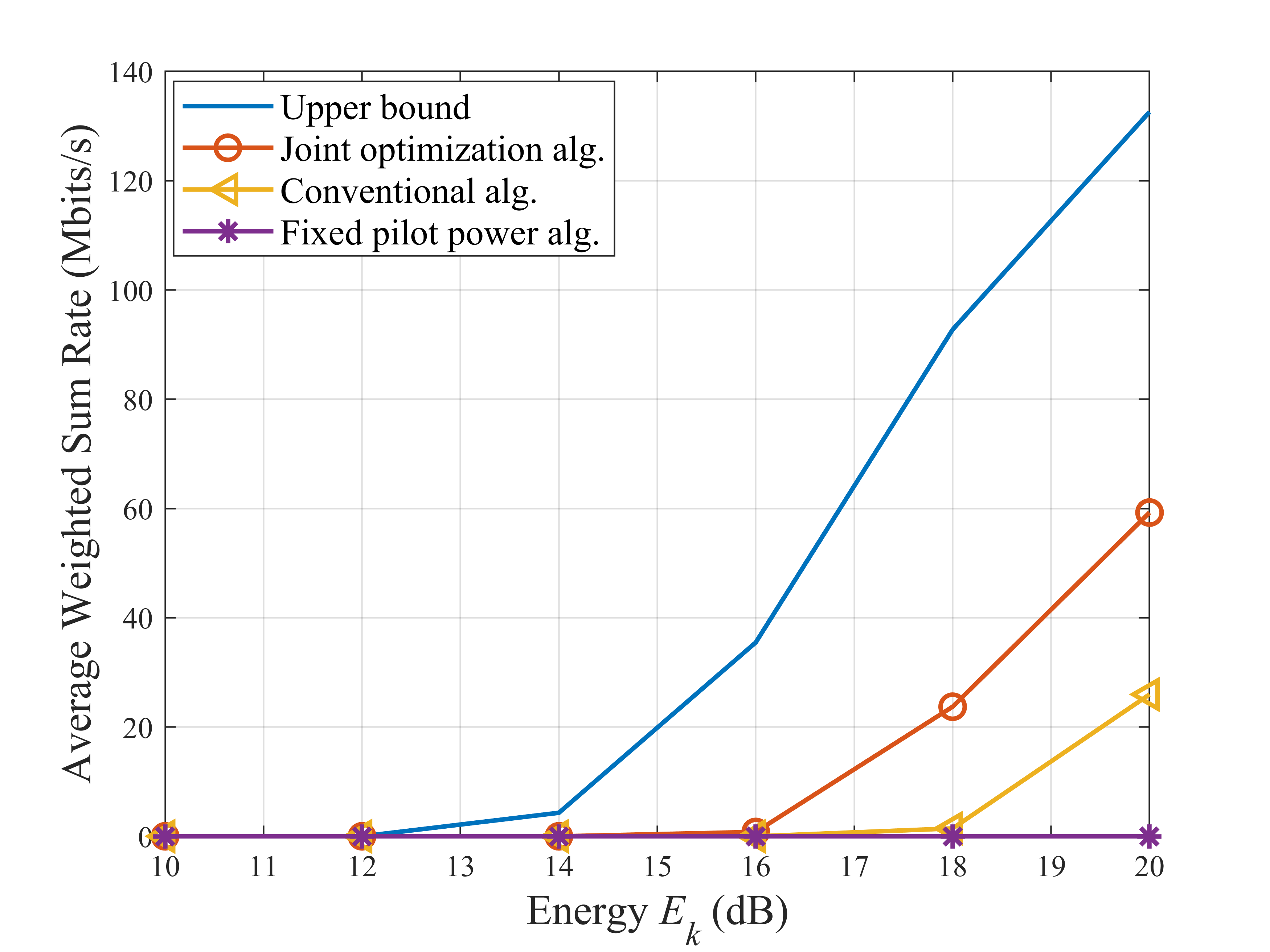}%
\caption*{(a) M=1}
\label{fig:MRC_performance_M1}}
\end{minipage}
\hfill
\begin{minipage}[t]{0.32\linewidth}
{\includegraphics[width=2.25in]{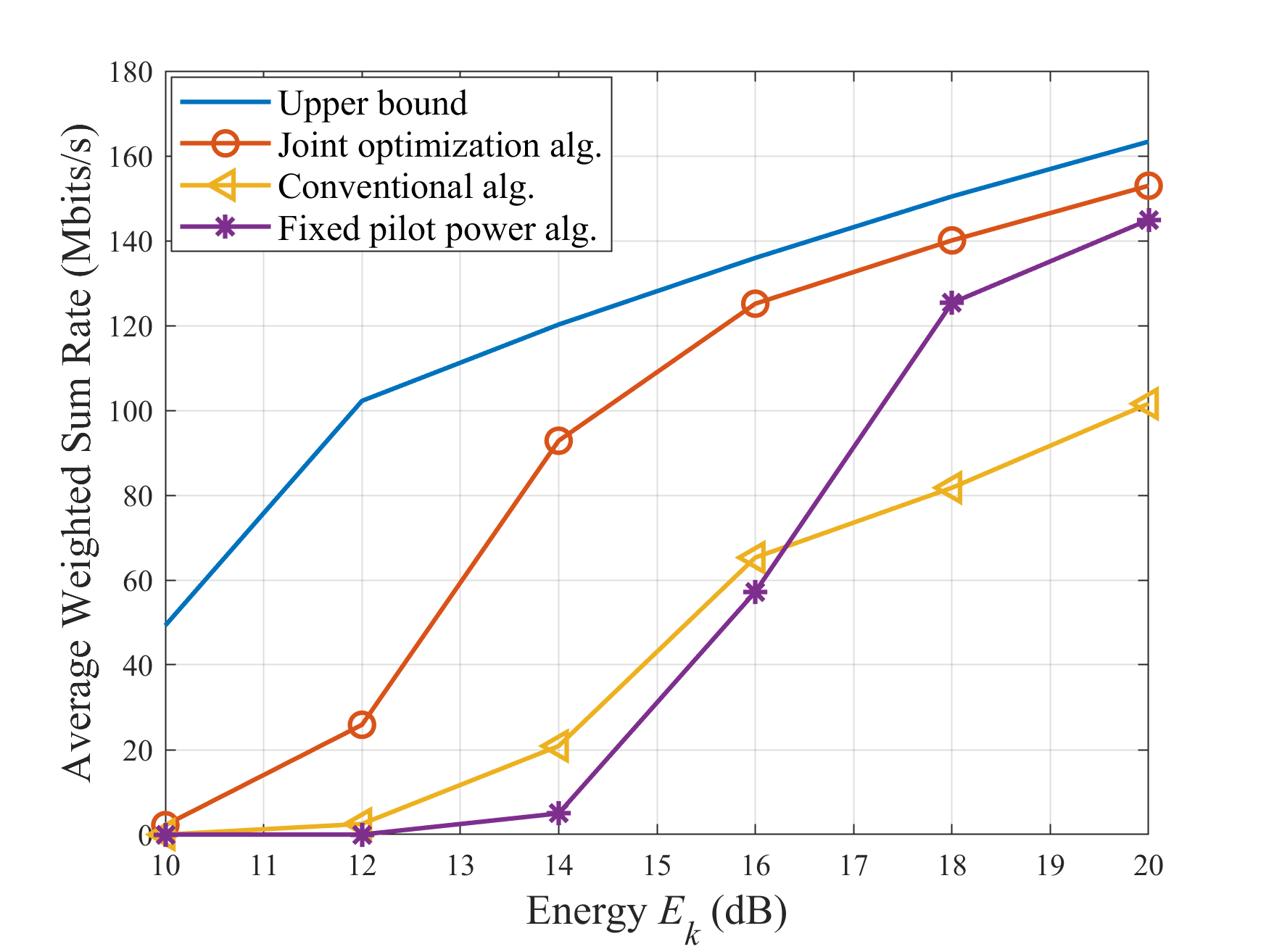}%
\label{fig:MRC_performance_M4}}
\caption*{(b) M=4}
\end{minipage}
\hfill
\begin{minipage}[t]{0.32\linewidth}
{\includegraphics[width=2.25in]{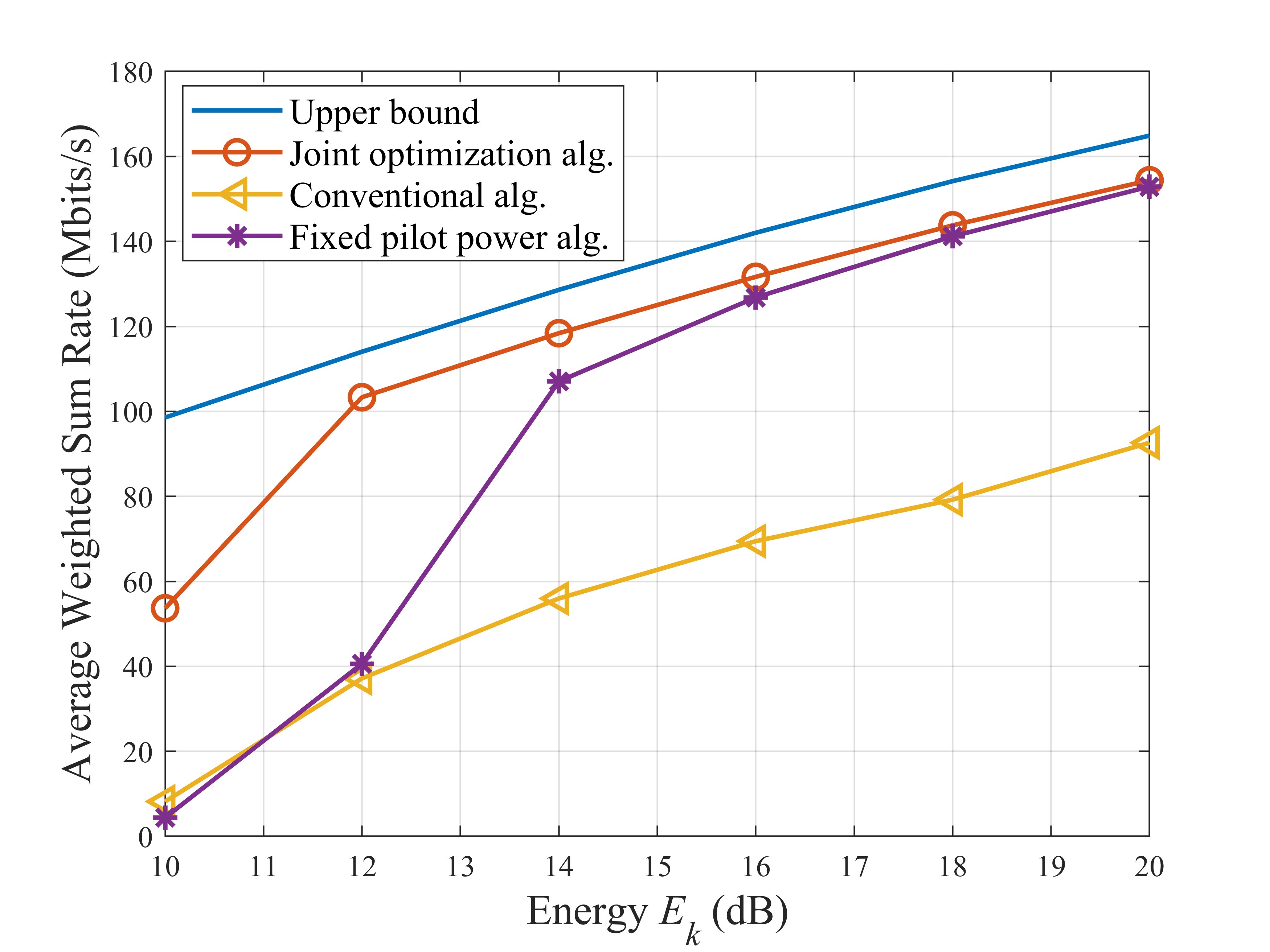}%
\label{fig:MRC_performance_M9}}
\caption*{(c) M=9}
\end{minipage}
\caption{Performance comparison with different numbers of APs for the MRC decoder.}
\label{fig_MRC_performance_comparison}
\end{figure*}

Next, the weighted sum rate for the FZF decoder is depicted in Fig. \ref{fig_FZF_performance_comparison}. As expected, our proposed algorithm can achieve the best performance. Particularly, similar to the MRC decoder, the fixed pilot power algorithm can meet the requirement of URLLC only when deploying more APs and using more energy. For conventional method, its performance could approach the upper bound with the increasing energy, which is different from the results of the MRC case. More importantly, an interesting observation is that $4$ APs provide a better performance than 9 APs, which reveals a tradeoff between deploying more APs and installing more antennas on each AP when adopting the FZF decoder.

\begin{figure*}
\begin{minipage}[t]{0.32\linewidth}
{\includegraphics[width=2.25in]{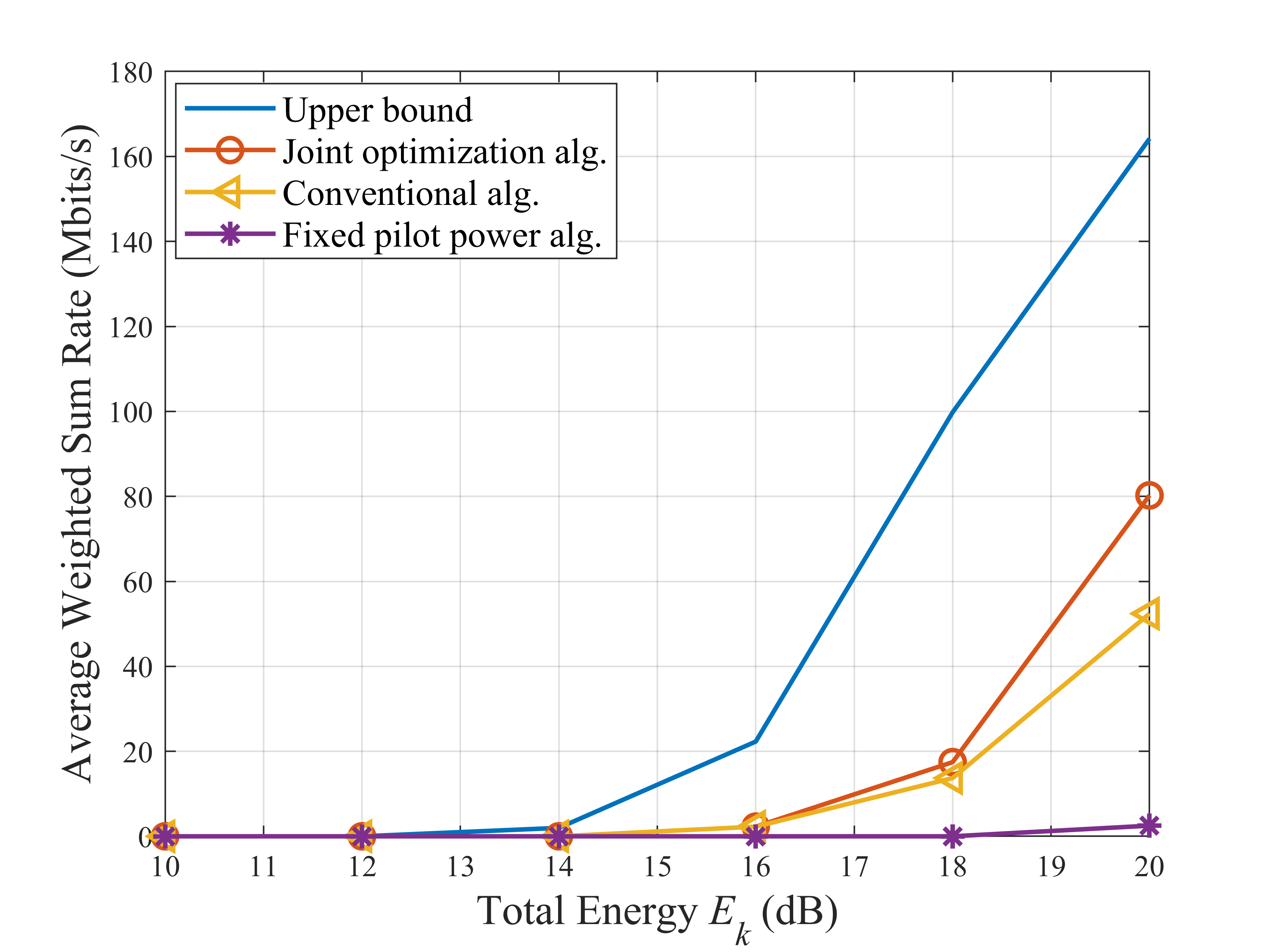}%
\caption*{(a) M=1}
\label{fig:FZF_performance_M1}}
\end{minipage}
\hfill
\begin{minipage}[t]{0.32\linewidth}
{\includegraphics[width=2.25in]{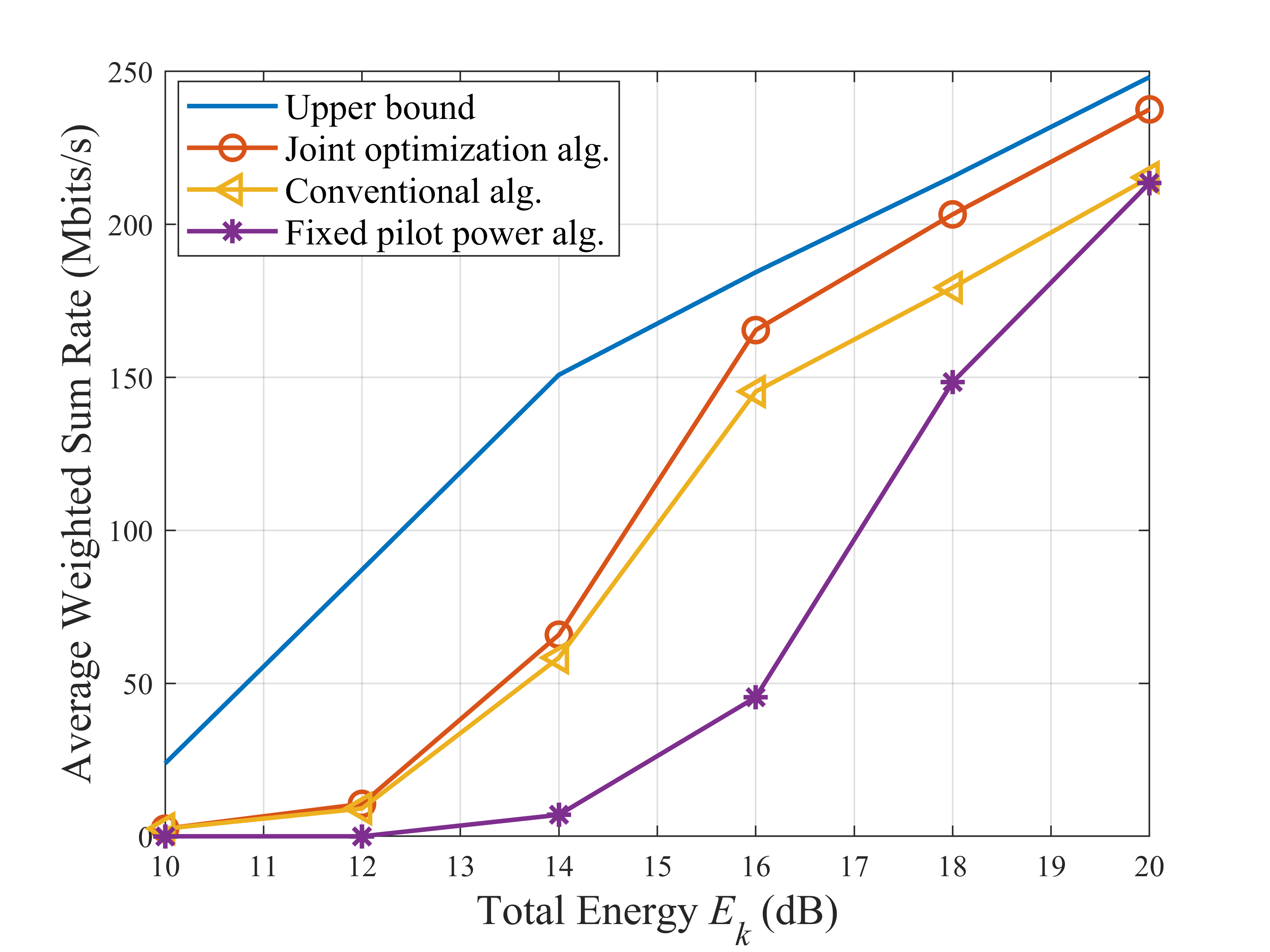}%
\label{fig:FZF_performance_M4}}
\caption*{(b) M=4}
\end{minipage}
\hfill
\begin{minipage}[t]{0.32\linewidth}
{\includegraphics[width=2.25in]{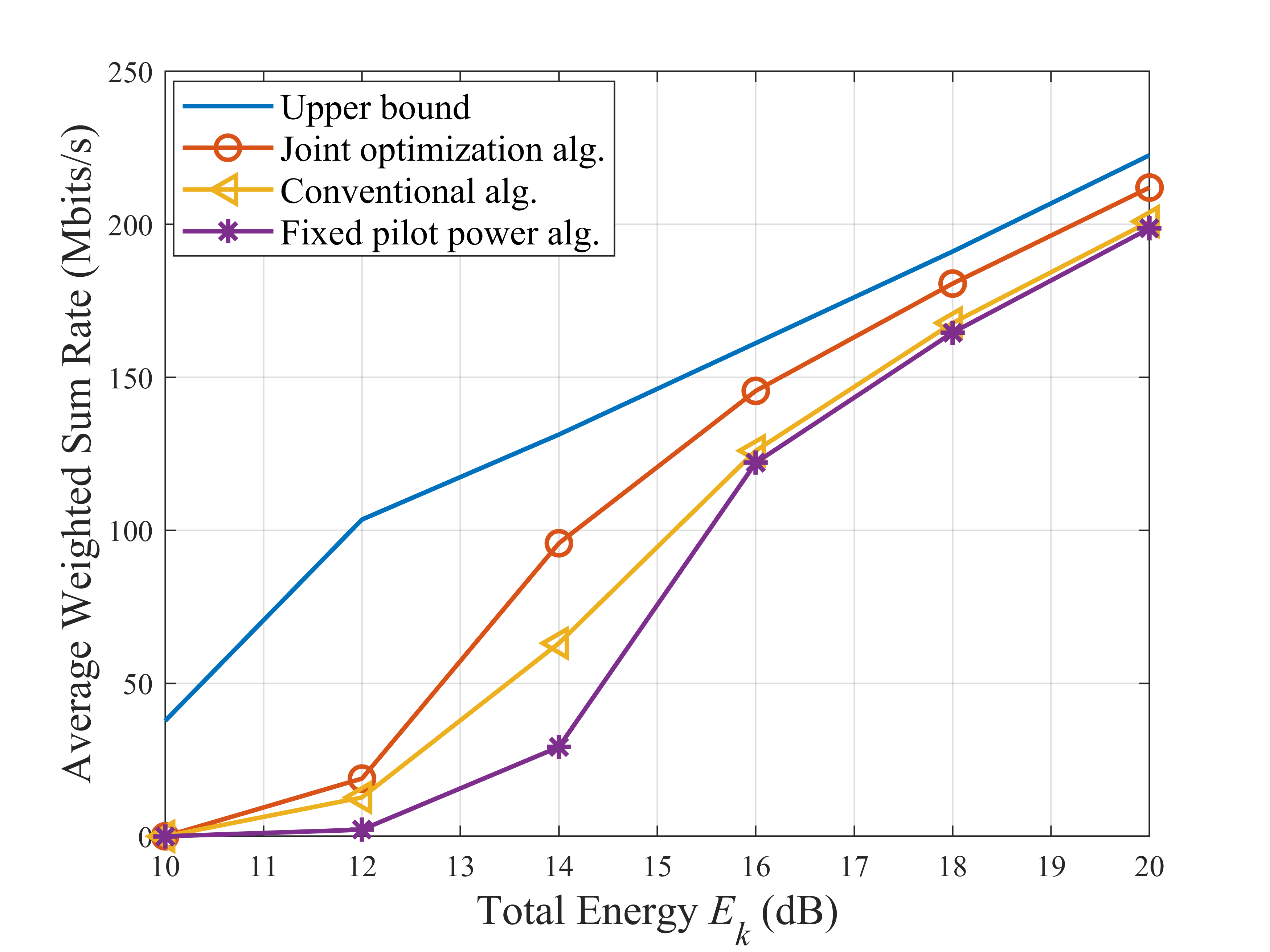}%
\label{fig:FZF_performance_M9}}
\caption*{(c) M=9}
\end{minipage}
\caption{Performance comparison with different numbers of APs for the FZF decoder.}
\label{fig_FZF_performance_comparison}
\end{figure*}

\subsection{Effect of Number of Devices}
In this subsection, we investigate the relationship between the number of devices and the system performance with the total energy of $E_k = 20$ ${\rm{dB}}$, $\forall k$. Here, we assume that the number of devices is smaller than the number of antennas deployed \textcolor{black}{at} each AP to ensure channel hardening. Fig. \ref{fig_MRC_vs_UEs} and Fig. 11 show the average performance versus the number of devices for MRC and FZF decoders, respectively. For \textcolor{black}{both} decoders, the performance of conventional method has an unpredictable trend, as the optimal solution based on \textcolor{black}{the} Shannon capacity \textcolor{black}{under IFBL} does not consider the penalty due to FBL. Besides, CF mMIMO can provide uniform services for all devices, while the centralized mMIMO cannot support URLLC for the devices that are far away from APs, leading to the fact that the system performance remains almost unchanged or even degraded with the increasing number of the devices for the MRC and FZF decoders, respectively. More importantly, CF mMIMO with the FZF decoder cannot provide URLLC for all devices when the number of devices $K$ approaches the number of each AP's antenna $N$.

\begin{figure*}
\begin{minipage}[t]{0.32\linewidth}
{\includegraphics[width=2.25in]{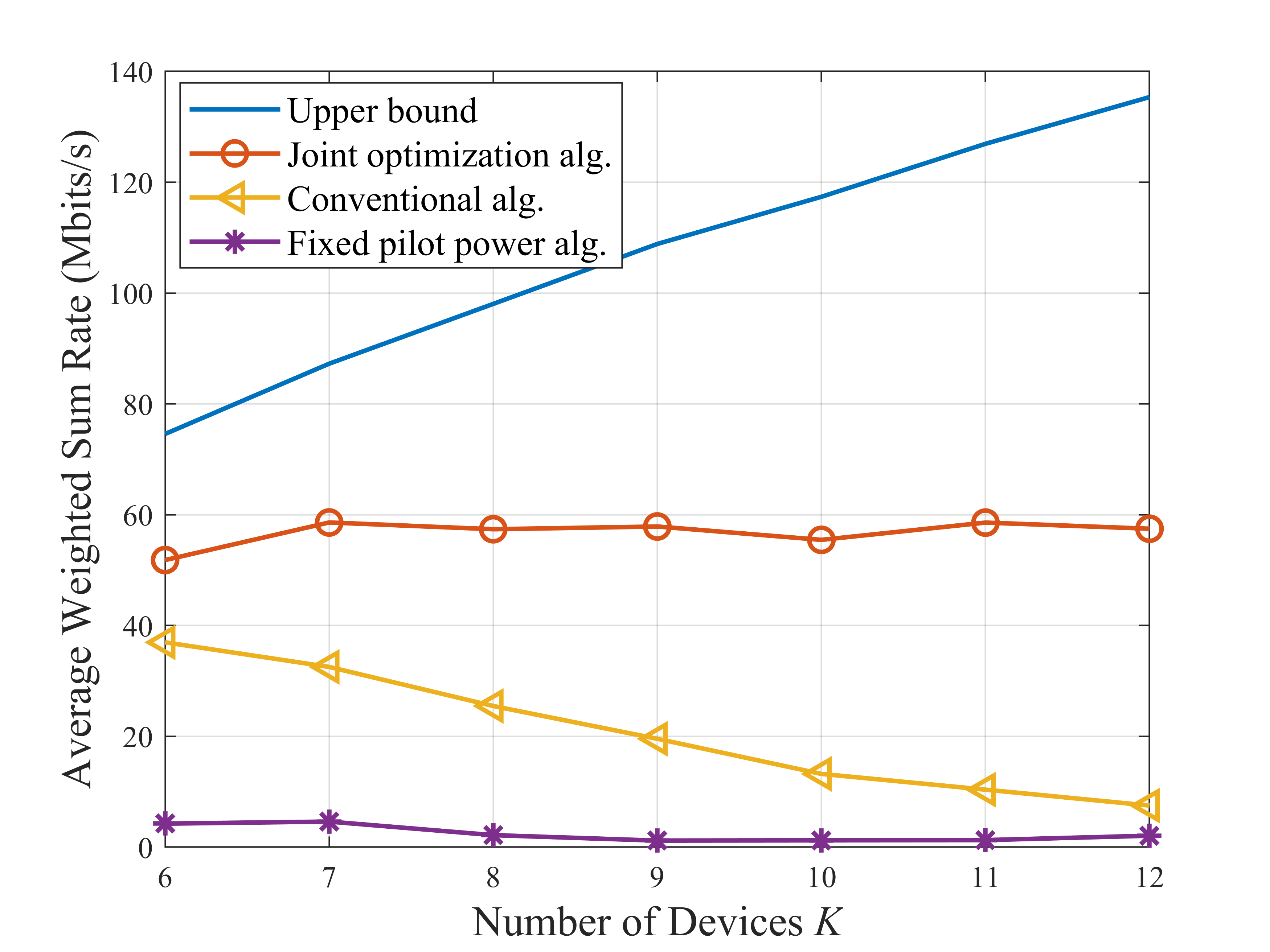}%
\caption*{(a) M=1}
\label{fig:MRC_vs_UEs_M1}}
\end{minipage}
\hfill
\begin{minipage}[t]{0.32\linewidth}
{\includegraphics[width=2.25in]{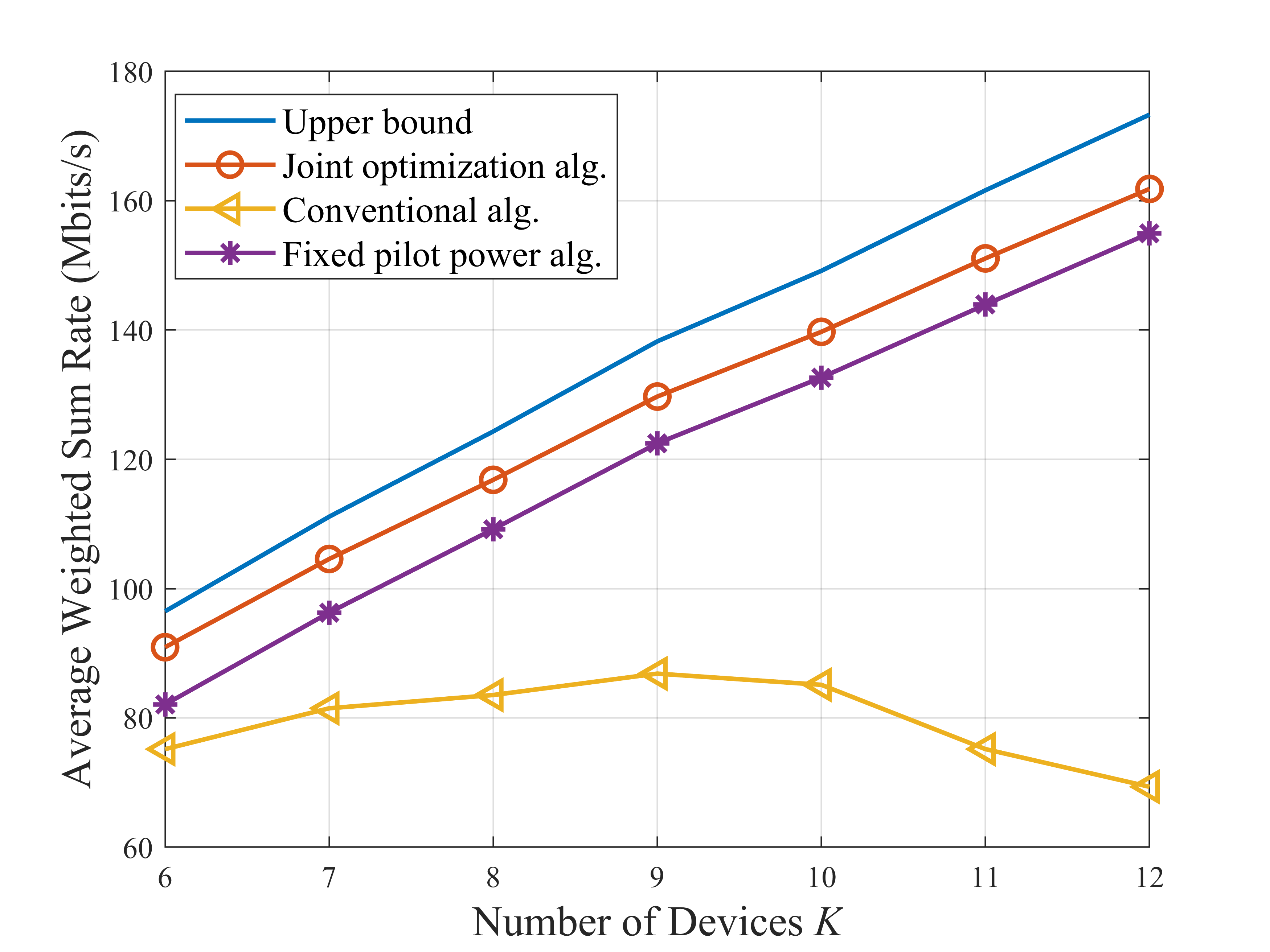}%
\label{fig:MRC_vs_UEs_M4}}
\caption*{(b) M=4}
\end{minipage}
\hfill
\begin{minipage}[t]{0.32\linewidth}
{\includegraphics[width=2.25in]{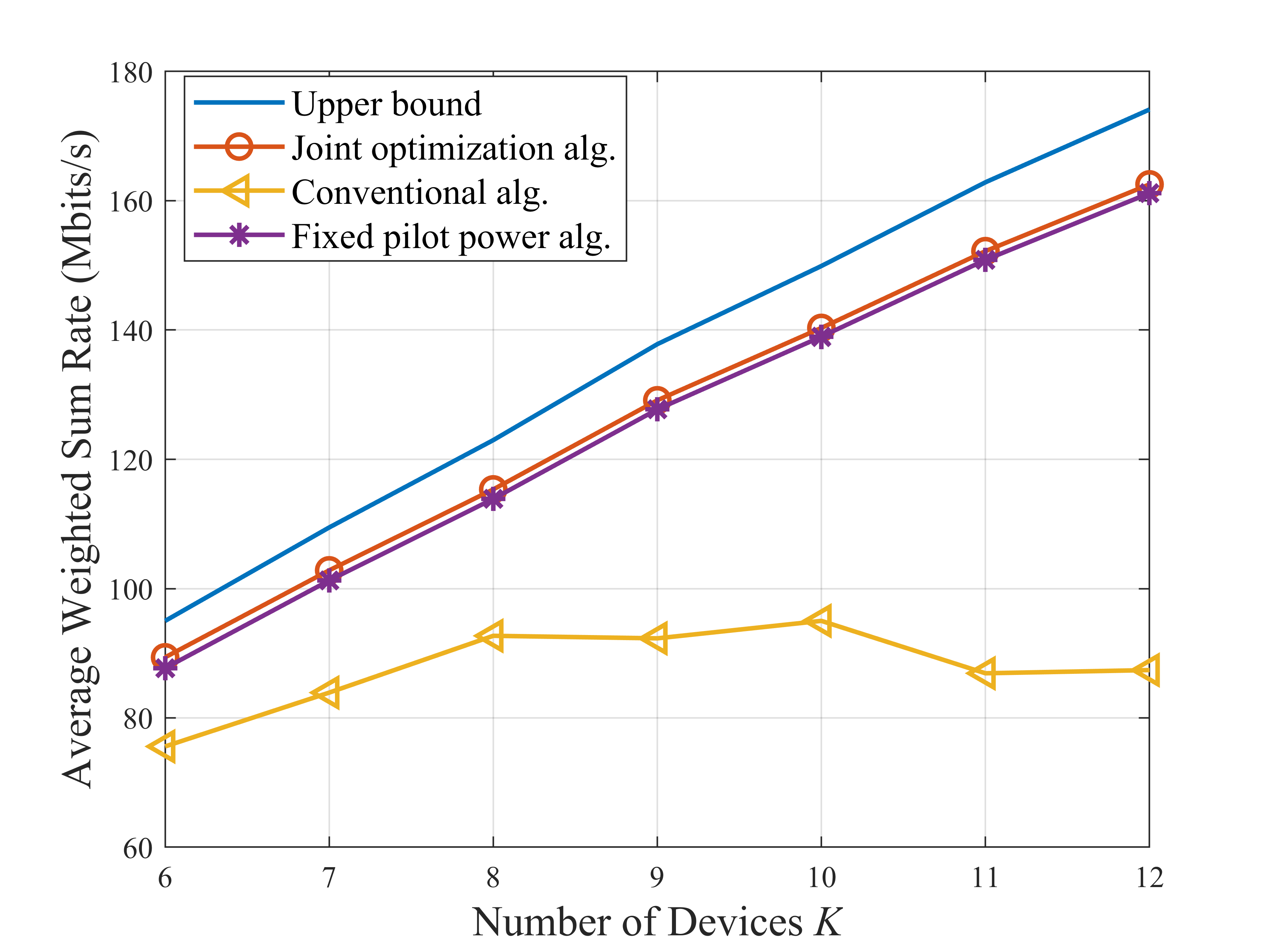}%
\label{fig:MRC_vs_UEs_M9}}
\caption*{(c) M=9}
\end{minipage}
\caption{Average Performance V.S. Number of devices for the MRC decoder. }
\label{fig_MRC_vs_UEs}
\end{figure*}

\begin{figure*}
\begin{minipage}[t]{0.32\linewidth}
{\includegraphics[width=2.25in]{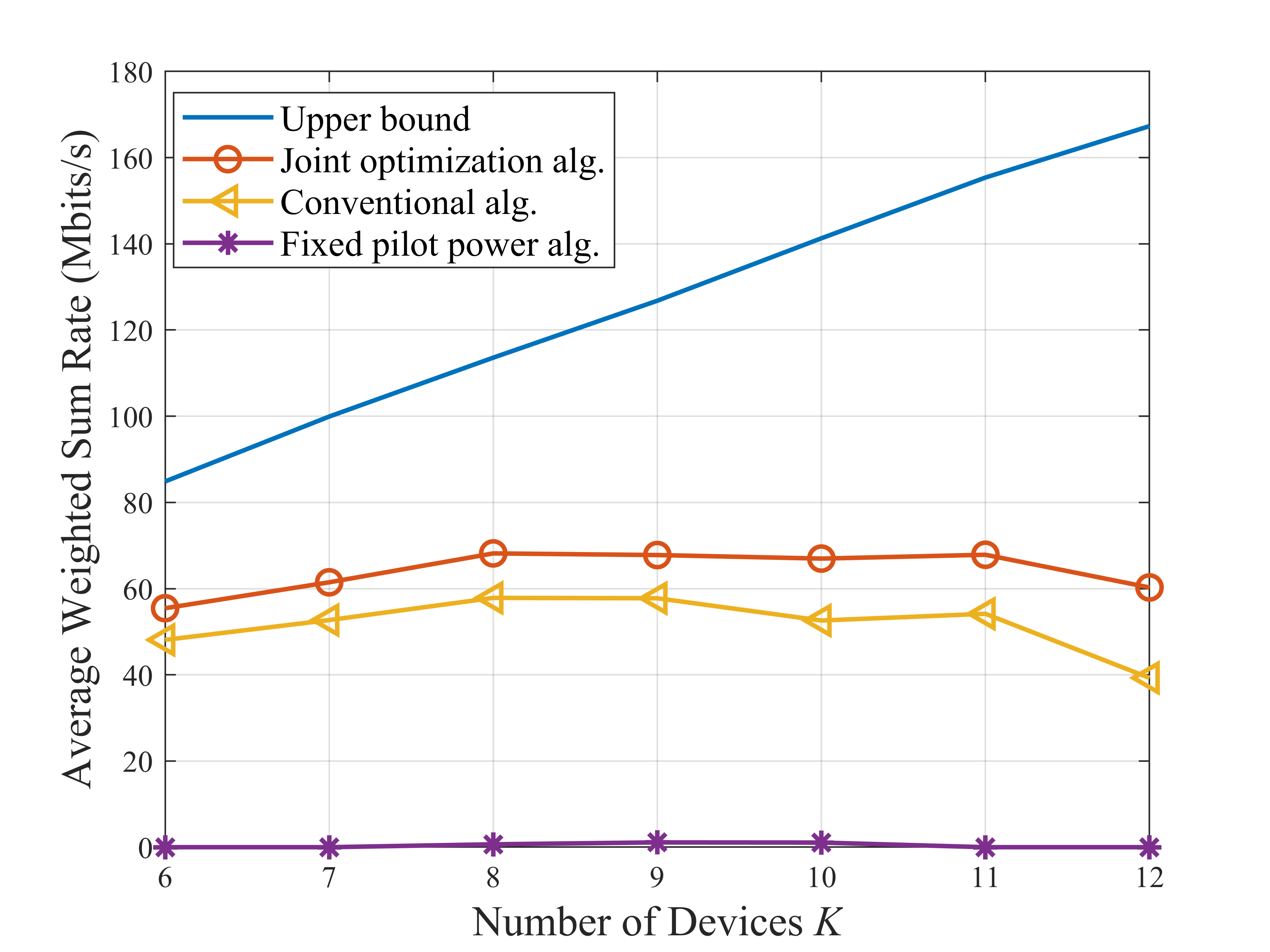}%
\caption*{(a) M=1}
\label{fig:FZF_vs_UEs_M1}}
\end{minipage}
\hfill
\begin{minipage}[t]{0.32\linewidth}
{\includegraphics[width=2.25in]{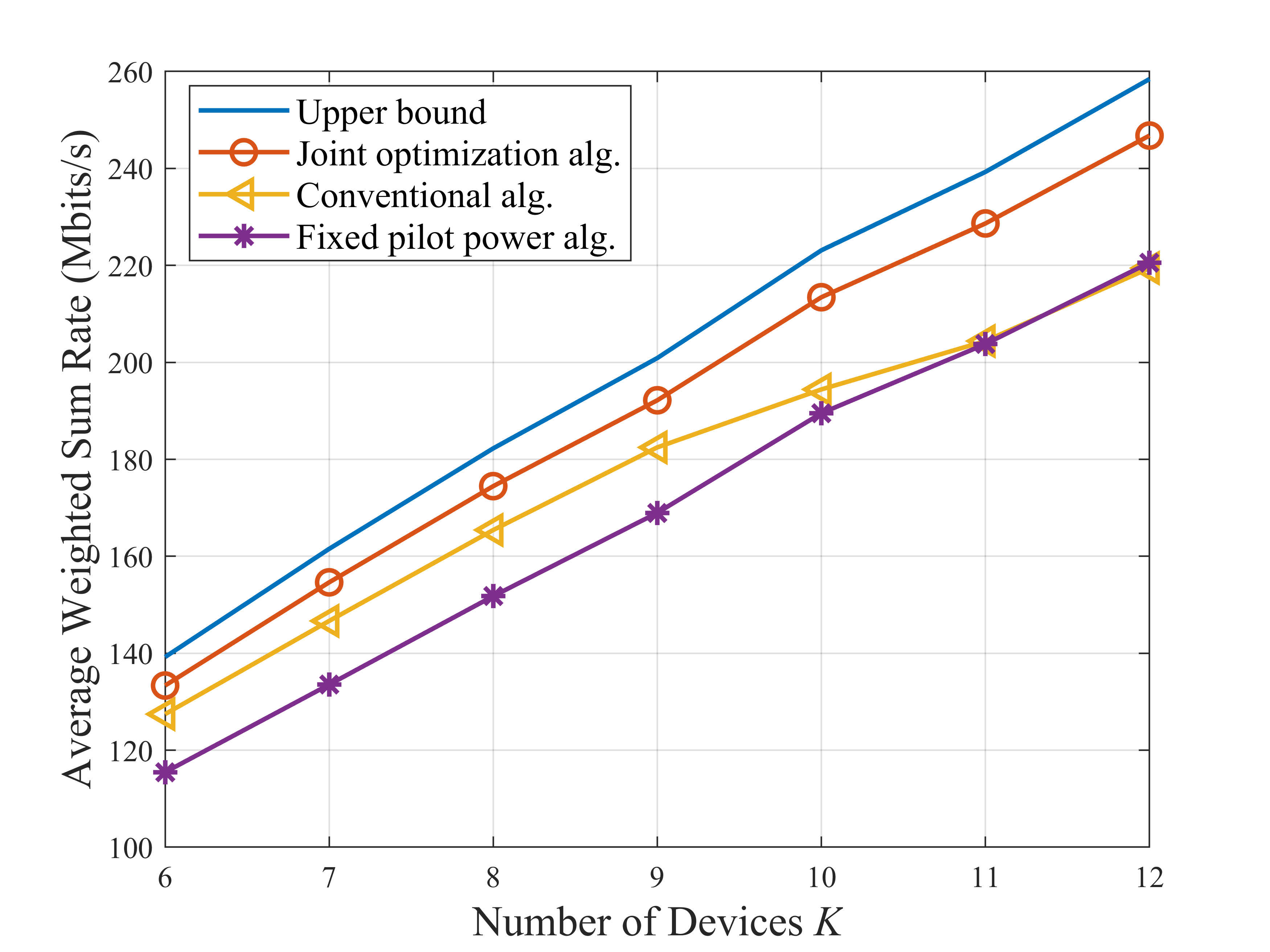}%
\label{fig:FZF_vs_UEs_M4}}
\caption*{(b) M=4}
\end{minipage}
\hfill
\begin{minipage}[t]{0.32\linewidth}
{\includegraphics[width=2.25in]{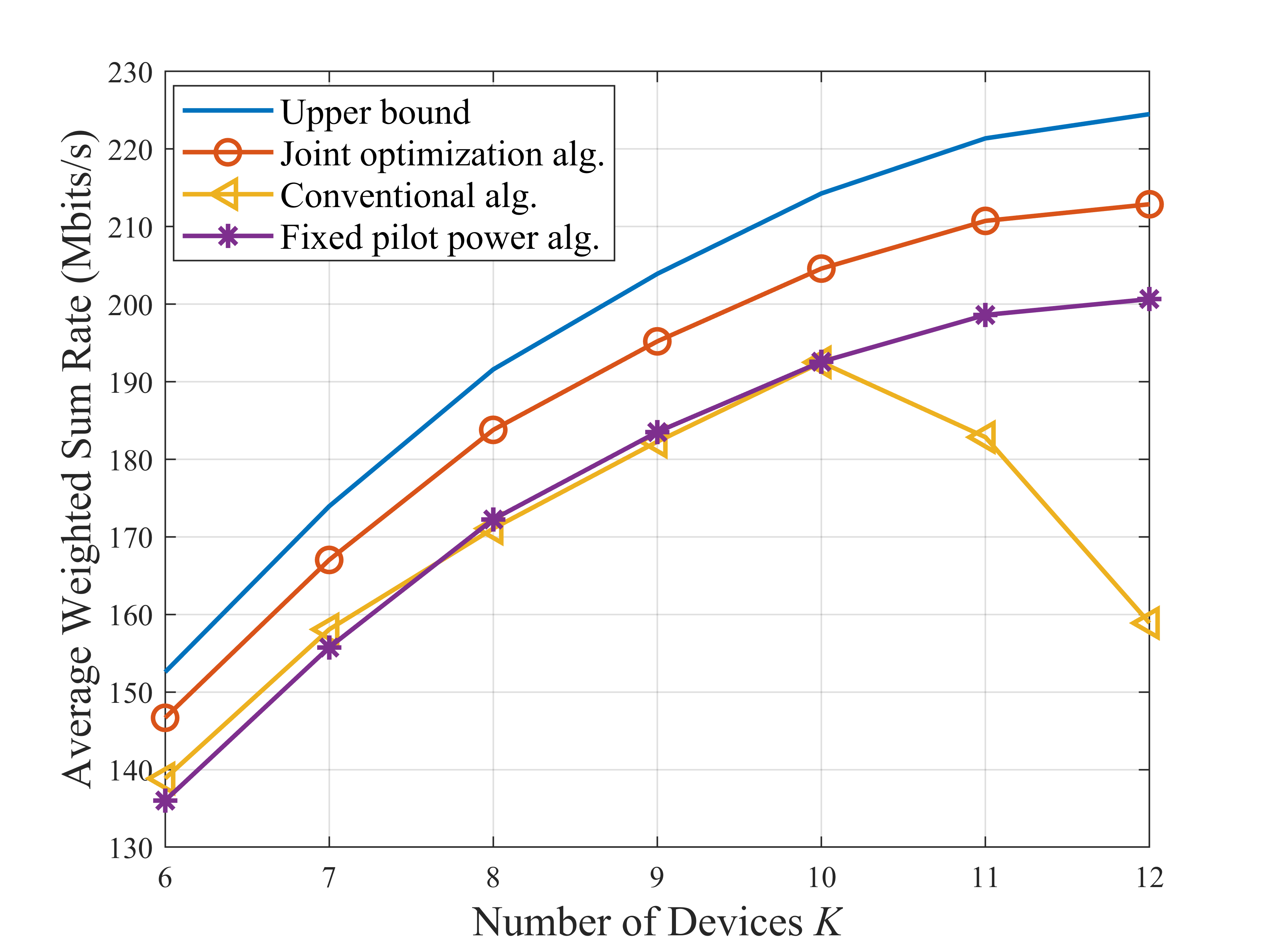}%
\label{fig:FZF_vs_UEs_M9}}
\caption*{(c) M=9}
\end{minipage}
\caption{Average Performance V.S. Number of devices for the FZF decoder. }
\label{fig_FZF_vs_UEs}
\end{figure*}

\section{Conclusion}
In this paper, we studied the resource allocation for uplink CF mMIMO systems to support URLLC in a smart factory. The closed-form LB data rates with imperfect CSI for both MRC and FZF decoders were derived, which is more tractable than the exact ergodic rate. Then, the joint optimization of the pilot and the payload power was proposed to maximize the weighted sum rate while considering limited energy, data rate, and DEP requirements. Finally, to tackle the non-convex problem, an iterative algorithm that uses SCA and GP was proposed. Simulation results demonstrated that the algorithm converges rapidly, and outperforms the existing benchmark algorithms for all cases, especially for devices with a lower energy budget, which demonstrates the effectiveness of our approach.

\begin{appendices}
\section{Proof of Theorem \ref{MRC_SINR_T}}
\label{MRC_SINR_P}
From (\ref{kth_SINR}), we need to derive the expressions of ${\left| {{\rm{DS}}_{k}} \right|^2}$, $\mathbb{E}\left( \left| {{\rm{LS}}_{k}} \right|^2 \right)$, $\mathbb{E}\left( \left| {{\rm{UI}}_{k,k'}} \right|^2 \right)$ and $\mathbb{E}\left( \left| {{\rm{N}}_{k}} \right|^2 \right)$, respectively.

We first compute ${\rm{DS}}_k$. Since $\hat {\bf{g}}_{m,k}$ and $\tilde {\bf{g}}_{m,k}$ are independent, we have
\begin{equation}
\setlength\abovedisplayskip{5pt}
\setlength\belowdisplayskip{5pt}
\label{MRC_DSk}
\begin{split}
	{\left| {{\rm{DS}}_{k}} \right|^2}   &= {\left| \mathbb{E}{\left\{ {\sum\limits_{m \in {\mathcal{M}}_k} {\sqrt {{p_k^d}} {{\left( {{\bf{\hat g}}_{m,k}} \right)}^H}\left( {{\bf{\hat g}}_{m,k} + {\bf{\tilde g}}_{m,k}} \right)} } \right\}} \right|^2}  \\
	 &= {N^2} {{p_k^d}} {\left( {\sum\limits_{m \in {\mathcal{M}}_k} {{\lambda _{m,k}}  } } \right)^2} .
\end{split}
\end{equation}

The term $\mathbb{E}\left( \left| {{\rm{LS}}_{k}} \right|^2 \right)$ is given by
\begin{equation}
\setlength\abovedisplayskip{5pt}
\setlength\belowdisplayskip{5pt}
\label{MRC_LSk}
\begin{split}
&\mathbb{E} \left\{ {{{\left| {{\rm{LS}}_{k}} \right|}^2}} \right\}   \\
= &{p_k^d} \mathbb{E} \left\{ {{{\left| {\sum\limits_{m \in {\mathcal{M}}_k}{{{\left( {{\bf{\hat g}}_{m,k}} \right)}^H}{\bf{g}}_{m,k} }  - N\sum\limits_{m \in {\mathcal{M}}_k}{ {\lambda _{m,k}}} } \right|}^2}} \right\} \\
 = &N{p_k^d}\sum\limits_{m \in {\mathcal{M}}_k} {{\lambda _{m,k}}{\beta _{m,k}}} .
 \end{split}
\end{equation}

Then, $\mathbb{E}\left( \left| {{\rm{UI}}_{k,k'}} \right|^2 \right)$ can be calculated as
\begin{equation}
\setlength\abovedisplayskip{5pt}
\setlength\belowdisplayskip{5pt}
\label{MRC_UIkk}
\begin{split}
 &\mathbb{E} \left( {{{\left| {{\rm{U}}{{\rm{I}}_{k,k'}}} \right|}^2}} \right)\\
  = &\mathbb{E} \left\{ {{{\left| {\sum\limits_{m \in {\mathcal{M}}_k} {\sqrt {p_{k'}^d} {{\left( {{{{\bf{\hat g}}}_{m,k}}} \right)}^H}{{\bf{g}}_{m,k'}}} } \right|}^2}} \right\}  \\
  = &\mathbb{E} \left\{ {{{\left| {\sum\limits_{m \in {\mathcal{M}}_k} {\sqrt {p_{k'}^d} {\hat{\alpha} _{m,k}}{{\left( {{{\bf{g}}_{m,k}} + {\bf{n}}_{m,k}^p} \right)}^H}{{\bf{g}}_{m,k'}}} } \right|}^2}} \right\}  \\
 = & p_{k'}^d \mathbb{E} \left\{ {{{\left| {\sum\limits_{m \in {\mathcal{M}}_k}{{\hat{\alpha} _{m,k}}{{\left( {{{\bf{g}}_{m,k}}} \right)}^H}{{\bf{g}}_{m,k'}}} } \right|}^2}}\right\}  \\
 & + p_{k'}^d \mathbb{E} \left\{{{{\left| {\sum\limits_{m \in {\mathcal{M}}_k} {{\hat{\alpha} _{m,k}}{{\left( {{\bf{n}}_{m,k}^p} \right)}^H}{{\bf{g}}_{m,k'}}} } \right|}^2}}\right\},
 \end{split}
\end{equation}
where ${\hat{\alpha} _{m,k}}$ is equal to ${\hat{\alpha} _{m,k}} = \frac{{\lambda}_{m,k}}{{\beta}_{m,k}}$. For each term in (\ref{MRC_UIkk}), we have
\begin{equation}
\setlength\abovedisplayskip{5pt}
\setlength\belowdisplayskip{5pt}
\label{UIkk_first}
\begin{split}
&\mathbb{E} \left\{ {{{\left| {\sum\limits_{m \in {\mathcal{M}}_k} {{\hat{\alpha} _{m,k}}{{\left( {{{\bf{g}}_{m,k}}} \right)}^H}{{\bf{g}}_{m,k'}}} } \right|}^2}} \right\} \\
=& \mathbb{E} \left\{ {\sum\limits_{m \in {\mathcal{M}}_k} {{{\left( {{\hat{\alpha} _{m,k}}} \right)}^2}{{\left( {{{\bf{g}}_{m,k}}} \right)}^H}{{\bf{g}}_{m,k'}}{{\left( {{{\bf{g}}_{m,k'}}} \right)}^H}{{\bf{g}}_{m,k}}} } \right\} \\
 = &N\sum\limits_{m \in {\mathcal{M}}_k}  {{{\left( {{\lambda _{m,k}}} \right)}^2}\frac{{{\beta _{m,k'}}}}{{{\beta _{m,k}}}}},
 \end{split}
\end{equation}
and
\begin{equation}
\setlength\abovedisplayskip{5pt}
\setlength\belowdisplayskip{5pt}
\label{UIkk_second}
\begin{split}
	&\mathbb{E} \left\{ {{{\left| {\sum\limits_{m \in {\mathcal{M}}_k} {{\hat{\alpha} _{m,k}}{{\left( {{\bf{n}}_{m,k}^p} \right)}^H}{{\bf{g}}_{m,k'}}} } \right|}^2}} \right\} \\
	=& \frac{N}{{Kp_{k'}^p}}\sum\limits_{m \in {\mathcal{M}}_k} {{{\left( {\frac{{{\lambda _{m,k}}}}{{{\beta _{m,k}}}}} \right)}^2}{\beta _{m,k'}}} .
\end{split}
\end{equation}

Finally, we compute $\mathbb{E}\left( \left| {{\rm{N}}_{k}} \right|^2 \right)$. Similar to (\ref{UIkk_second}), we have
\begin{equation}
\setlength\abovedisplayskip{5pt}
\setlength\belowdisplayskip{5pt}
\label{MRC_Nk}
\mathbb{E} \left( {{{\left| {{{\rm{N}}_k}} \right|}^2}} \right) = N \sum\limits_{m \in {\mathcal{M}}_k}  {{\lambda _{m,k}}}.
\end{equation}

Substituting (\ref{MRC_DSk}), (\ref{MRC_LSk}), (\ref{UIkk_first}), and (\ref{UIkk_second}) into (\ref{kth_SINR}), we obtain $\hat \gamma _k^{{\rm{MRC}}}$ in (\ref{MRC_SINR_LB}).

\section{Proof of Theorem \ref{FZF_SINR_T}}
\label{FZF_SINR_P}
Similar to the MRC decoder, we need to derive the expressions of ${\left| {{\rm{DS}}_{k}} \right|^2}$, $\mathbb{E}\left( \left| {{\rm{LS}}_{k}} \right|^2 \right)$, $\mathbb{E}\left( \left| {{\rm{UI}}_{k,k'}} \right|^2 \right)$ and $\mathbb{E}\left( \left| {{\rm{N}}_{k}} \right|^2 \right)$, respectively.

Before deriving ${\left| {{\rm{DS}}_{k}} \right|^2}$, we need to calculate \textcolor{black}{the decoding} vector for the FZF decoder. The coefficient of normalized vector can be derived as
\begin{equation}
\setlength\abovedisplayskip{5pt}
\setlength\belowdisplayskip{5pt}
\label{Normalized_ZF}
{{{\mathbb{E}} \left\{ {{{\left\| {{{{\bf{\hat G}}}_m}{{\left[ {{\bf{\hat G}}_m^H{{{\bf{\hat G}}}_m}} \right]}^{ - 1}}{{\bf{e}}_k}} \right\|}^2}}\right\} }}=\frac{1}{{\left( {N - K} \right){\lambda _{m,k}}}}.
\end{equation}

Then, ${\left| {{\rm{DS}}_{k}} \right|}$ can be derived as
\begin{equation}
\setlength\abovedisplayskip{5pt}
\setlength\belowdisplayskip{5pt}
\label{DS_ZF}
\begin{split}
 \left| {{\rm{DS}}_{k}} \right| & =\mathbb{E} \left\{ {\sum\limits_{m \in {\mathcal{M}}_k} {{{\left( {{\bf{ a}}_{m,k}} \right)}^H}{\bf{g}}_{m,k} \sqrt {{p_k^d}} } } \right\}  \\
  &= \sqrt {{p_k^d}\left( {N - K} \right)} \sum\limits_{m \in {\mathcal{M}}_k} {\sqrt {{\lambda _{m,k}}} } .
\end{split}
\end{equation}

Next, the leakage power can be formulated as
\begin{equation}
\setlength\abovedisplayskip{5pt}
\setlength\belowdisplayskip{5pt}
\label{LS_ZF}
\begin{split}
	\mathbb{E} \left\{ {{{\left| {{\rm{LS}}_{k}} \right|}^2}} \right\} &= \mathbb{E} \left\{ {{{\left| {\sum\limits_{m \in {\mathcal{M}}_k} {{{\left( {{\bf{ a}}_{m,k}} \right)}^H}{\bf{g}}_{m,k}} \sqrt {{p_k^d}}  - {{\rm{DS}}_{k}} } \right|}^2}} \right\} \\ 
	&= {p_k^d}\sum\limits_{m \in {\mathcal{M}}_k} {\left( {{\beta _{m,k}} - {\lambda _{m,k}}} \right)} .
\end{split}
\end{equation}

The term $\mathbb{E}\left( \left| {{\rm{UI}}_{k,k'}} \right|^2 \right)$ can be expressed as
\begin{equation}
\setlength\abovedisplayskip{5pt}
\setlength\belowdisplayskip{5pt}
\label{UI_ZF}
\begin{split}
&\mathbb{E} \left\{ {{{\left| {{\rm{UI}}_{k,k'}} \right|}^2}} \right\} \\
= &\mathbb{E} \left\{ {{{\left| {\sum\limits_{m \in {\mathcal{M}}_k} {{{\left( {{\bf{ a}}_{m,k}} \right)}^H}{\bf{g}}_{m,k'}\sqrt {{p_{k'}^d}} } } \right|}^2}} \right\}  \\
= & {p_{k'}^d}\left( {N - K} \right) \mathbb{E}\! \left\{\! {\sum\limits_{m \in {\mathcal{M}}_k} \!\!\!\!{{\lambda _{m,k}}\!\left( {{\beta _{m,k'}} - {\lambda _{m,k'}}} \right)\!\!\left[ {{\bf{\hat G}}_m^H{{{\bf{\hat G}}}_m}} \right]_{k,k}^{ - 1}}\!\! } \right\}  \\
= &{p_{k'}^d}\sum\limits_{m \in {\mathcal{M}}_k} {\left( {{\beta _{m,k'}} - {\lambda _{m,k'}}} \right)} .
\end{split}
\end{equation}

The noise term can be derived as
\begin{equation}
\setlength\abovedisplayskip{5pt}
\setlength\belowdisplayskip{5pt}
\label{Noise_ZF}
\mathbb{E} \left\{ {{{\left| {{N_k}} \right|}^2}} \right\} =  \mathbb{E}  \left\{ {{{\left\| {\sum\limits_{m \in {\mathcal{M}}_k} {{{\left( {{\bf{ a}}_{m,k}} \right)}}} } \right\|}^2}} \right\} = |{\cal M}_k|,
\end{equation}
where $|{\cal M}_k|$ means the cardinality of the set ${\cal M}_k$.

Finally, we complete the proof by substituting the expressions of (\ref{DS_ZF}), (\ref{LS_ZF}), (\ref{UI_ZF}), and (\ref{Noise_ZF}) into the SINR expression.

\section{Proof of Lemma \ref{lnx}}
\label{proof_lnx}
The inequality in (\ref{lemma1}) can be readily proved by substituting the expressions of $\rho$ and $\delta$ into (\ref{lemma1}). Then, we define $J\left( x \right) = \ln \left( {1 + x} \right) - \rho \ln x - \delta$, the first-order derivative is given by
\begin{equation}
\setlength\abovedisplayskip{5pt}
\setlength\belowdisplayskip{5pt}
\label{J_derivative}
\frac{{dJ\left( x \right)}}{{dx}} = \frac{{x - \rho \left( {1 + x} \right)}}{{\left( {1 + x} \right)x}} = \frac{{x\left( {1 + \hat x} \right) - \hat x \left( {1 + x} \right)}}{{\left( {1 + \hat x} \right)\left( {1 + x} \right)x}}.
\end{equation}

Since both $x$ and $\hat x$ are positive values, the sign of $\frac{{dJ\left( x \right)}}{{dx}}$ only depends on the numerator. Let us define $H(x) = {x\left( {1 + \hat x} \right) - \hat x \left( {1 + x} \right)}$, and then the first-order derivative of $H(x)$ is given by $H'(x) = 1$, which means $H(x)$ is monotonically increasing. Consequently, we have $H(x) \ge 0$ when $x \ge \hat x$ since $H\left (\hat x\right) = 0$, which indicates that $J\left( x \right)$ is a monotonically increasing function if $x \ge \hat x$. Similarly, we can prove the $J\left( x \right)$ is a monotonically decreasing function if $x < \hat x$. As a result, we complete the proof by showing that $J\left( x\right)$ is always larger than $J\left( \hat x\right) = 0$,

\section{Proof of Lemma \ref{lemma_MRC_Sub}}
\label{proof_MRC_Substitution}
Using (\ref{MRC_sinr_theta}) and (\ref{MRC_sinr_sigma}), we have
\begin{equation}
\setlength\abovedisplayskip{5pt}
\setlength\belowdisplayskip{5pt}
 \label{theta_sigma}
 \begin{split}
 	\frac{{{\theta _k}}}{{{\sigma _k}}} &= \frac{{\sum\limits_{m \in {\mathcal{M}}_k}  {\left[ {Kp_k^p{{\left( {{\beta _{m,k}}} \right)}^2}\prod\limits_{n \ne m} {\left( {Kp_k^p{\beta _{n,k}} + 1} \right)} } \right]} }}{{\prod\limits_{m \in {\mathcal{M}}_k} {\left( {Kp_k^p{\beta _{m,k}} + 1} \right)} }} \\
 	& = \sum\limits_{m \in {\mathcal{M}}_k} {{\lambda _{m,k}}} ,
 \end{split} 
\end{equation}
where $\lambda_{m,k}$ is given in (\ref{gama_mk}).

Using (\ref{MRC_sinr_sigma}) and (\ref{MRC_sinr_xi}), we have
\begin{equation}
\setlength\abovedisplayskip{5pt}
\setlength\belowdisplayskip{5pt}
\label{xi_sigma}
\begin{split}
	 \frac{{{{\xi}_{k,k'}}}}{{{\sigma _k}}} &= \frac{{\sum\limits_{m \in {\mathcal{M}}_k} {\left[ {Kp_k^p{{\left( {{\beta _{m,k}}} \right)}^2}{\beta _{m,k'}}\prod\limits_{n \ne m} {\left( {Kp_k^p{\beta _{n,k}} + 1} \right)} } \right]} }}{{\prod\limits_{m \in {\mathcal{M}}_k} {\left( {Kp_k^p{\beta _{m,k}} + 1} \right)} }} \\ 
	 &= \sum\limits_{m \in {\mathcal{M}}_k} {{\lambda _{m,k}}{\beta _{m,k'}}} .
\end{split}
\end{equation}

Substituting (\ref{xi_sigma}) and (\ref{theta_sigma}) into (\ref{MRC_SINR_LB}), we have
\begin{equation}
\setlength\abovedisplayskip{5pt}
\setlength\belowdisplayskip{5pt}
\label{MRC_SINR_trans}
\hat \gamma_k^{\rm{MRC}}= \frac{{Np_{k}^d{{\left( {{\theta _k}} \right)}^2}}}{{\sum\limits_{k' = 1}^K {{p_{k'}^d}} {\sigma _k}{\xi _{k,k'}} + {\sigma _k}{\theta _k}}}.
\end{equation}

\section{Proof of Theorem \ref{theta_T}}
\label{proof_theta_T}
By taking the logarithm operator for the left hand side of (\ref{MRC_theta_LB}), we have
\begin{equation}
\setlength\abovedisplayskip{5pt}
\setlength\belowdisplayskip{5pt}
\label{Fx}
\begin{split}
	  &\ln \left( {{\theta _k}} \right)  
	  \\= &\ln \left( {K {{\rm{e}}^x}} \right) + \ln \left( {\sum\limits_{m \in {\mathcal{M}}_k} {\left[ {{{\left( {{\beta _{m,k}}} \right)}^2}\prod\limits_{n \ne m} {\left( {K {\beta _{n,k}}{{\rm{e}}^x} + 1} \right)} } \right]} } \right) \\
	   \triangleq& F\left( x \right), 
\end{split}
\end{equation}
where $x$ is given by $x = \ln \left(p_k^p \right)$.

It is readily found that (\ref{Fx}) can be transformed into the following equivalent form
\begin{equation}
\setlength\abovedisplayskip{5pt}
\setlength\belowdisplayskip{5pt}
\label{rw_Fx}
F\left( x \right) = \ln \left( {K {{\rm{e}}^x}} \right)  + \ln \left( {\sum\limits_{m \in {\mathcal{M}}_k} {\left[ {{\upsilon _m}{{\rm{e}}^{\Gamma_m x}}} \right]} } \right),
\end{equation}
where $\upsilon _m$ and $\Gamma_m$ are certain positive constant values. Note that the expressions of $\upsilon _m$ and $\Gamma_m$ are not needed since we only need to prove that (\ref{rw_Fx}) is a convex function of $x$. Then, by using Jensen's inequality, we have
\begin{equation}
\setlength\abovedisplayskip{5pt}
\setlength\belowdisplayskip{5pt}
\label{In_Fx}
F\left( x \right) \ge {a_k}x + {c_k},
\end{equation}
where $a_k$ and $c_k$ are given in (\ref{a_k}) and (\ref{c_k}), respectively.

Finally, we complete the proof by taking the exponential operation for both sides of (\ref{In_Fx}) and using $x = \ln \left(p_k^p\right)$.

\section{Proof of Lemma \ref{lemma_FZF_Substitution}}
\label{proof_FZF_Substitution}
By using ${\varpi _k}$ in (\ref{ZF_varpi}) and $\vartheta _{k,k}$ in (\ref{ZF_vartheta}), we have
\begin{equation}
\setlength\abovedisplayskip{5pt}
\setlength\belowdisplayskip{5pt}
\label{ZF_verify1}
\begin{split}
	\frac{{{\varpi _k}}}{{{\vartheta _{k,k}}}} &= \frac{{\sum\limits_{m \in {\mathcal{M}}_k} {\left[ {\sqrt {K{p^p_{k}}{{\left( {{\beta _{m,k}}} \right)}^2}} \prod\limits_{n \ne m} {\sqrt {K{p^p_{k}}{\beta _{n,k}} + 1} } } \right]} }}{{\prod\limits_{m \in {\mathcal{M}}_k} {\sqrt {K{p^p_{k}}{\beta _{m,k}} + 1} } }}\\
	& = {\sum\limits_{m \in {\mathcal{M}}_k} {\sqrt {{\lambda _{m,k}}} } } .
\end{split}
\end{equation}

By using (\ref{ZF_vartheta}) and (\ref{ZF_mu}), we have
\begin{equation}
\setlength\abovedisplayskip{5pt}
\setlength\belowdisplayskip{5pt}
\label{ZF_verify2}
\begin{split}
	\frac{{{\mu _{k,k'}}}}{{{{\left( {{\vartheta _{k,k'}}} \right)}^2}}} &=\frac{{\sum\limits_{m \in {\mathcal{M}}_k}  {\left[ {{\beta _{m,k'}}\prod\limits_{n \ne m} {\left( {K{p^p_{k'}}{\beta _{n,k'}} + 1} \right)} } \right]} }}{{\prod\limits_{m \in {\mathcal{M}}_k} {\left( {K{p^p_{k'}}{\beta _{m,k'}} + 1} \right)} }}  \\
	&= {\sum\limits_{m \in {\mathcal{M}}_k} {\left( {{\beta _{m,k'}} - {\lambda_{m,k'}}} \right)} } .
\end{split}
\end{equation}

Finally, we complete the proof by substituting (\ref{ZF_verify1}) and (\ref{ZF_verify2}) into the SINR expression for the FZF case in (\ref{SINR_ZF}).

\section{Proof of Theorem \ref{varpi_T}}
\label{proof_varpi_T}
The term $\ln \left[ {{{\left( {{\varpi _k}} \right)}^2}\prod\limits_{k' \ne k}^K {{{\left( {{\vartheta _{k,k'}}} \right)}^2}} } \right]$ can be expressed as
\begin{equation}
\setlength\abovedisplayskip{5pt}
\setlength\belowdisplayskip{5pt}
\label{ZF_Fk}
 2\ln \left( {{\varpi _k}} \right) + \sum\limits_{k' \ne k}^K {\ln \left( {{{\left( {{\vartheta _{k,k'}}} \right)}^2}} \right)} \triangleq  {D_k}\left( {{{\bf{x}}}} \right),
\end{equation}
where ${\bf{x}}$ is equal to ${\bf{x}} = \left[x_1, x_2,\cdot \cdot \cdot, x_K  \right]$ with $x_k = \ln \left(p_k^p\right)$.

Next, we need to derive the second-order derivatives of $\ln \left({\varpi _k}\right)$ and $\ln \left({\vartheta _{k,k'}}\right)$  to prove that they are both convex functions. We substitute $x_{k} = \ln \left(p_k^p\right)$ into $\ln \left({\varpi _k}\right)$, denoted as
\begin{equation}
\setlength\abovedisplayskip{5pt}
\setlength\belowdisplayskip{5pt}
 \label{ZF_ln_varpi}
\begin{split}
& \ln \left(\left( {{\varpi _k}} \right)^2 \right)  \\
 = & \ln \left[ \left( \sum\limits_{m \in {\mathcal{M}}_k}{ {\sqrt {K{{\rm{e}}^{x_{k}}}{{\left( {{\beta _{m,k}}} \right)}^2}} \prod\limits_{n \ne m} {\sqrt {K{{\rm{e}}^{x_{k}}}{\beta _{n,k}} + 1} } } } \right)^2 \right]  \\
= & \ln\left(K{{\rm{e}}^{x_{k}}} \right) \! + \!\ln \left[ \!\left( \sum\limits_{m \in {\mathcal{M}}_k}{ {{{\beta _{m,k}}} \prod\limits_{n \ne m} {\sqrt {K{{\rm{e}}^{x_{k}}}{\beta _{n,k}} + 1} } } } \right)^2\! \right]  \\
= & \ln\left(K{{\rm{e}}^{x_{k}}} \right) \!+\! 2 \ln \left(\sum\limits_{m \in {\mathcal{M}}_k}  {{\hat \upsilon _{m}}{{\rm{e}}^{{\hat \Gamma_{m}} x_k}}}  \right) ,
\end{split}
\end{equation}
where ${\hat \upsilon _{m}}$ and $\hat \Gamma_{m}$ are certain positive constant values. Similar to the proof of Theorem \ref{theta_T}, the expressions of ${\hat \upsilon _{m}}$ and $\hat \Gamma_{m}$ are not needed since we only need to prove that (\ref{ZF_ln_varpi}) is a convex function of $x_k$. Similar to $F\left(x\right)$ in (\ref{rw_Fx}), it is readily to prove $\ln \left(\left( {{\varpi _k}} \right)^2 \right)$ is a convex function.

The second-order derivative of $\ln \left( {{{\left( {{\vartheta _{k,k'}}} \right)}^2}} \right)$ is given by
\begin{equation}
\setlength\abovedisplayskip{5pt}
\setlength\belowdisplayskip{5pt}
\label{ZF_ln_vartheta}
\frac{{{\partial ^2}\ln \left( {{{\left( {{\vartheta _{k,k'}}} \right)}^2}} \right)}}{{\partial {{\left( {{x_{k'}}} \right)}^2}}} = \frac{{\sum\limits_{m \in {{\cal M}_k}} {\left( {K{\beta _{m,k'}}{{\rm{e}}^{{x_{k'}}}}} \right)} }}{{\sum\limits_{m \in {{\cal M}_k}} {\left( {K{\beta _{m,k'}}{{\rm{e}}^{{x_{k'}}}} + 1} \right)} }} > 0,
\end{equation}
where $x_{k'}$ is equal to $x_{k'} = \ln \left(p_{k'}^p\right)$. Then, $\ln \left( {\vartheta _{k,k'}}\right)$ is a convex function of $x_{k'}$.

Then, it is readily to calculate $\frac{{{\partial ^2}\ln \left( {{{\left( {{\vartheta _{k,k'}}} \right)}^2}} \right)}}{{\partial \left( {{x_{k'}}} \right)\partial \left( {{x_k}} \right)}} = 0$ and $\frac{{{\partial ^2}\ln \left(\left( {{\varpi _k}} \right)^2 \right)}}{{\partial \left( {{x_{k'}}} \right)\partial \left( {{x_k}} \right)}} = 0$, and thus the Hessian matrix of $D_k\left(\bf {x} \right)$ in (\ref{ZF_Fk}) is positive semi-definite. As a result, by using Jensen's inequality, we have
\begin{equation}
\setlength\abovedisplayskip{5pt}
\setlength\belowdisplayskip{5pt}
\label{ZF_Dk}
{D_k}\left( {\bf{x}} \right) \ge \sum\limits_{j = 1}^K {{b^k_j}{x_j}}  + {d_k},
\end{equation}
where $b^k_j$ and $d_k$ are given in (\ref{ZF_bk}) and (\ref{ZF_Ck}), respectively.

Finally, we complete this proof by taking exponential operation for both sides of (\ref{ZF_Dk}).

\end{appendices}

\bibliographystyle{IEEEtran}
\bibliography{myref}

\end{document}